\documentclass{elsarticle}

\usepackage{epsfig}
\usepackage{subcaption}
\usepackage{calc}
\usepackage{amssymb}
\usepackage{amstext}
\usepackage{amsmath}
\usepackage{amsthm}
\usepackage{multicol}
\usepackage{pslatex}
\usepackage{graphicx}
\usepackage{listings, xcolor}
\usepackage{colortbl}
\usepackage{flushend}
\usepackage[hyphens]{url}
\usepackage{tcolorbox}

\usepackage{pgfplots}
\usepackage{tikz}
\definecolor{bblue}{HTML}{4F81BD}
\definecolor{dblue}{HTML}{000066}
\definecolor{rred}{HTML}{C0504D}
\definecolor{ggreen}{HTML}{9BBB59}
\definecolor{ppurple}{HTML}{9F4C7C}
\definecolor{yyellow}{HTML}{BDAD11}
\definecolor{mmaroon}{HTML}{800000}

\definecolor{verylightgray}{rgb}{.97,.97,.97}
\definecolor{codegreen}{rgb}{0,0.6,0}
\definecolor{codegray}{rgb}{0.5,0.5,0.5}
\definecolor{codepurple}{rgb}{0.58,0,0.82}
\definecolor{backcolour}{rgb}{0.9,0.9,0.9}

\lstdefinelanguage{Solidity}{
	keywords=[1]{anonymous, assembly, assert, balance, break, call, callcode, case, catch, class, constant, continue, constructor, contract, debugger, default, delegatecall, delete, do, else, emit, event, experimental, export, external, false, finally, for, function, gas, if, implements, import, in, indexed, instanceof, interface, internal, is, length, library, log0, log1, log2, log3, log4, memory, modifier, new, payable, pragma, private, protected, public, pure, push, require, return, returns, revert, selfdestruct, send, solidity, storage, struct, suicide, super, switch, then, this, throw, transfer, true, try, typeof, using, value, view, while, with, addmod, ecrecover, keccak256, mulmod, ripemd160, sha256, sha3, sub, swap1, pop, dup2, gt}, 
	keywordstyle=[1]\color{blue}\bfseries,
	keywords=[2]{address, bool, byte, bytes, bytes1, bytes2, bytes3, bytes4, bytes5, bytes6, bytes7, bytes8, bytes9, bytes10, bytes11, bytes12, bytes13, bytes14, bytes15, bytes16, bytes17, bytes18, bytes19, bytes20, bytes21, bytes22, bytes23, bytes24, bytes25, bytes26, bytes27, bytes28, bytes29, bytes30, bytes31, bytes32, enum, int, int8, int16, int24, int32, int40, int48, int56, int64, int72, int80, int88, int96, int104, int112, int120, int128, int136, int144, int152, int160, int168, int176, int184, int192, int200, int208, int216, int224, int232, int240, int248, int256, mapping, string, uint, uint8, uint16, uint24, uint32, uint40, uint48, uint56, uint64, uint72, uint80, uint88, uint96, uint104, uint112, uint120, uint128, uint136, uint144, uint152, uint160, uint168, uint176, uint184, uint192, uint200, uint208, uint216, uint224, uint232, uint240, uint248, uint256, var, void, ether, finney, szabo, wei, days, hours, minutes, seconds, weeks, years},	
	keywordstyle=[2]\color{blue}\bfseries,
	keywords=[3]{block, blockhash, coinbase, difficulty, gaslimit, number, timestamp, msg, data, gas, sender, sig, value, now, tx, gasprice, origin},	
	keywordstyle=[3]\color{violet}\bfseries,
	identifierstyle=\color{black},
	sensitive=false,
	comment=[l]{//},
	morecomment=[s]{/*}{*/},
	commentstyle=\color{black}\ttfamily,
	stringstyle=\color{red}\ttfamily,
	morestring=[b]',
	morestring=[b]"
}

\lstdefinestyle{style1}{
    backgroundcolor=\color{backcolour},   
    commentstyle=\color{codegreen},
    keywordstyle=\color{blue},
    numberstyle=\tiny\color{black},
    stringstyle=\color{codepurple},
    basicstyle=\footnotesize,
    breakatwhitespace=false,         
    breaklines=true,                 
    captionpos=t,                    
    keepspaces=true,                 
    numbers=left,                    
    numbersep=1pt,                  
    showspaces=false,                
    showstringspaces=false,
    showtabs=false,                  
    tabsize=2
}

\usepackage[ligature, inference]{semantic}
\usepackage{caption}

\theoremstyle{definition}
\newtheorem{definition}{Definition}[section]
\theoremstyle{definition}
\newtheorem{example}{Example}[section]
\theoremstyle{definition}

\theoremstyle{definition}

\theoremstyle{definition}

\theoremstyle{definition}

\makeatletter
\def\ps@pprintTitle{%
 \let\@oddhead\@empty
 \let\@evenhead\@empty
 \def\@oddfoot{}%
 \let\@evenfoot\@oddfoot}
\makeatother

\begin{document}
\begin{frontmatter}
\title{EtherClue: Digital investigation of attacks on Ethereum smart contracts}

\author[um]{Simon Joseph Aquilina}
\ead{simon-joseph.aquilina.15@um.edu.mt}
\author[unipi,arc]{Fran Casino}
 \ead{francasino@unipi.gr}
\author[um]{Mark Vella\corref{cor1}}
\ead{mark.vella@um.edu.mt}
\cortext[cor1]{Corresponding author}
\author[um,umdlt]{Joshua Ellul}
\ead{joshua.ellul@um.edu.mt}
\author[unipi,arc]{Constantinos Patsakis}
\ead{kpatsak@unipi.gr}
\address[um]{Department of Computer Science, University of Malta, Msida, Malta}
\address[umdlt]{Centre for Distributed Ledger Technologies, University of Malta, Msida, Malta}
\address[unipi]{Department of Informatics, University of Piraeus, 80 Karaoli \& Dimitriou str., 18534 Piraeus, Greece}
 \address[arc]{Information Management Systems Institute, Athena Research Center, Artemidos 6, Marousi 15125, Greece}
\journal{Blockchain: Research and Applications}
\begin{abstract}
Programming errors in Ethereum smart contracts can result in catastrophic financial losses from stolen cryptocurrency. While vulnerability detectors can prevent vulnerable contracts from being deployed, this does not mean that such contracts will not be deployed. Once a vulnerable contract is instantiated on the blockchain and becomes the target of attacks, the identification of exploit transactions becomes indispensable in assessing whether it has been actually exploited and identifying which malicious or subverted accounts were involved. 

In this work, we study the problem of post-factum investigation of Ethereum attacks using Indicators of Compromise (IoCs) specially crafted for use in the blockchain. IoC definitions need to capture the side-effects of successful exploitation in the context of the Ethereum blockchain. Therefore, we define a model for smart contract execution, comprising multiple abstraction levels that mirror the multiple views of code execution on a blockchain. Subsequently, we compare IoCs defined across the different levels in terms of their effectiveness and practicality through \emph{EtherClue}, a prototype tool for investigating Ethereum security incidents. Our results illustrate that coarse-grained IoCs defined over blocks of transactions can detect exploit transactions with less computation; however, they are contract-specific and suffer from false negatives. On the other hand, fine-grained IoCs defined over virtual machine instructions can avoid these pitfalls at the expense of increased computation which are nevertheless applicable for practical use.
\end{abstract}
\end{frontmatter}

\section{Introduction}
\label{sec:introduction}

Nowadays, Ethereum is the largest public blockchain supporting smart contract execution. While introducing flexibility in terms of programmable transactions, as defined by Turing complete programming languages, the consequence is an enlarged attack surface \cite{atzei2017survey,chen2020survey}. The immediate venue for attacks is presented by the smart contract code itself whenever it deviates from the intended functionality. This problem is aggravated by incomplete knowledge of how a smart contract is executed at runtime by inexperienced programmers. Security vulnerabilities at this level can be similar to those found in non-blockchain code, e.g. arithmetic overflows, reentrancy, and poor randomness. Others are specific to the particular nature of smart contracts, e.g. call to unknown third-party contract code, improperly checking the return status of external calls, and denial of service (DoS) vulnerabilities related to gas-based mitigation of infinite execution bugs or call-stack depth limits, among many others. Furthermore, the number of currency/token-based contracts deployed on Ethereum aggravate the classic vulnerability of insufficiently sanitised user-controlled inputs.

Other entry points for attacks are presented by the runtime environment supporting contract code execution. Ethereum's infrastructure requires a virtual machine for code execution, while the consensus protocol's complexity is increased as a result of taking into account contract execution during block mining. The case of malicious miners being selective of which transactions, or in which order, they are included inside a mined block, is one such example. By doing so, an attacker can exploit time or transaction order-dependent logic inside smart contracts. Finally, Ethereum's external environment comprises code that interacts with Ethereum nodes, e.g. dApps and off-chain crypto wallets, further extend the attack surface \cite{dappattacks}. At this level, attacks exploit vectors within the technologies in question, e.g. DOM-based XSS for dApps, or spyware in the case of desktop/mobile applications. Ultimately, these attacks can tamper with the submitted transactions in various ways.

In recent years, we have witnessed a number of notorious security incidents concerning the Ethereum blockchain that resulted in significant financial losses or drastic measures being taken to avoid consequences. The DAO crowd-funding platform incident is perhaps one of the most representative of these, as on June 18th, 2016, attackers managed to transfer \$60M under their control \cite{DAO}. The funds were salvaged only through a hard fork, a measure which goes directly against the principle of allowing autonomous authorities to be arranged foregoing central authority \cite{o2017smart}. In April 2018, an integer overflow vulnerability was the root cause for the attack on BECToken, resulting in the theft of $10^{58}$ tokens. Just earlier this year, a potential multi-million dollar disaster for users of Defi Saver was only averted by the timely action of its development team when notified of a vulnerability in one of their digital asset management contracts \cite{DefiSaver}.

This work aims to identify exploit transactions and, therefore, the focus is on vulnerabilities that can be exploited solely via maliciously crafted transactions. These are submitted either directly from an externally owned account, i.e. an account not associated with code, or indirectly through a malicious smart contract. Ideally, such vulnerabilities would have been detected prior to smart contract instance creation, especially when one considers blockchain immutability. Therefore, tools that perform static analysis of smart contract code, identifying security vulnerabilities inside the source code (e.g. Solidity) or decompiled Ethereum Virtual Machine (EVM) bytecode of smart contracts, comprise the most common approaches. Notable examples include Oyente \cite{luu2016making}, analysis tools based on Gigahorse \cite{grech2019gigahorse}, as well as EtherTrust \cite{grieco2020echidna}, Securify  \cite{tsankov2018securify}, Zeus \cite{kalra2018zeus} and the dynamic verification tool for absence of reentrancy vulnerabilities \cite{grossman2017online}. Others take a testing approach and focus on generating test inputs using a fuzzing approach \cite{jiang2018contractfuzzer,grieco2020echidna,liu2018reguard}, comparing test outcomes with test oracles. However, in both approaches, comprehensive execution path coverage remains an open challenge. As a result, vulnerable contracts may still be deployed on the blockchain. In such cases, exploit transactions can be identified \textit{post-factum} during a subsequent investigation with the help of Digital Forensics and Incident Response (DFIR) tools. Investigations of this nature can help, first and foremost, to understand whether the smart contract in question has been exploited. In the affirmative case, the identification of exploit transactions provides links to the source accounts. These accounts may either belong to the perpetrators or victims whose accounts have been compromised. Furthermore, the linked accounts may be associated with further smart contracts, which at this point would have to be considered malicious, with any future transactions originating from them considered suspicious.

Indicators of Compromise (IoC) constitute evidence of intrusion and are typically directly related to the long-term side-effects of successfully executing malware \cite{kao2018dynamic}. For example, in the case of WannaCry ransomware attacks on a Windows machine, investigators can identify which artifact variant is in play, and therefore which recovery method could be effective, by checking whether the values \texttt{cmd.exe /c} and \texttt{C:\textbackslash ProgramData\textbackslash aucdehyopp032\textbackslash tasksche.exe} can be found under the \sloppy{\texttt{"HKLM\textbackslash System\textbackslash CurrentControlSet\textbackslash Services\textbackslash mssecsvc2.0"}} key used for persistence. Another IoC can be based on the auto-propagation code that exploits EternalBlue \cite{wannacry}. This step is observable within a network packet capture in the form of SMB traffic attempting a connection to \texttt{IPC\$} and followed by an unusually large \texttt{NT Trans} SMB request. When one compares these host and network-level IoCs to Ethereum's state transitions, it is evident that we must consider radically different approaches.

We address the problem of responding to attacks on Ethereum smart contracts akin to responding to computer intrusions. In fact, during the earlier stages of the incident response process, a triage operation typically identifies compromised devices. This is a live forensics operation conducted while machines are still powered on. Indicators of Compromise (IoC) constitute forensic evidence of the intrusion and are the key elements of triage representing long-term residue of attack steps \cite{kao2018dynamic}. While similar IoCs could have been considered for attacks targeting Ethereum's operational environment, this approach does not apply at the smart contract level. If we consider contract execution steps as a state machine, we would rather be dealing with state transitions that update Ethereum blocks, add transactions to mined blocks and manipulate the EVM's internal state. Therefore, our first undertaking is precisely abstracting smart contracts execution in a model that considers all three levels. 

\begin{figure}[!ht]
  \centering
   \includegraphics[page=1,trim = 0mm 0mm 0mm 25mm, clip, width=.8\linewidth]{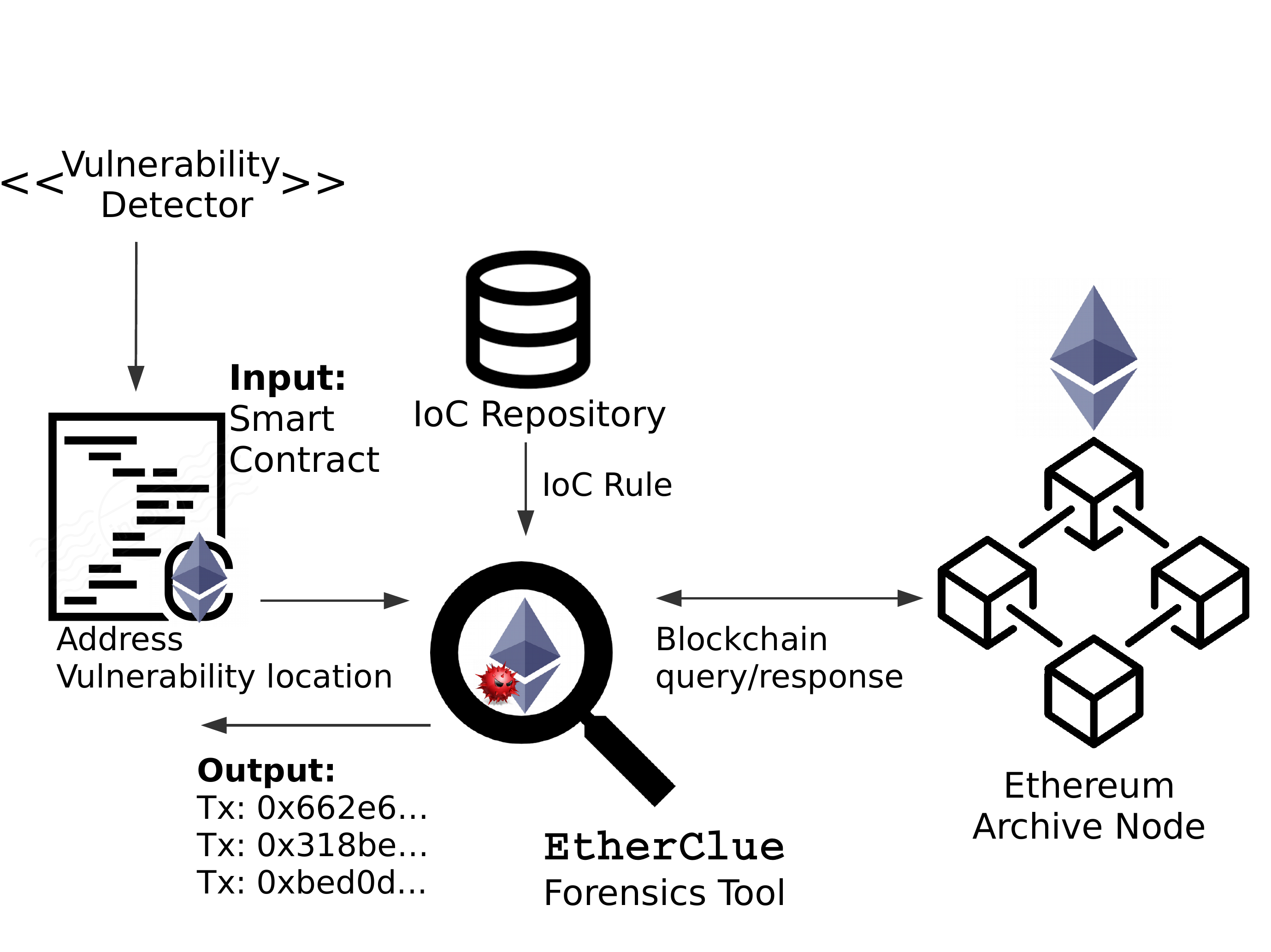}
  \caption{EtherClue DFIR tool.}
  \label{fig:etherclue}
\end{figure}

The result is the \textit{EtherClue} DFIR tool (Figure \ref{fig:etherclue}), which given the address of a contract account and a vulnerable code location, it makes use of an IoC repository to identify potential exploit transactions from within an archive node that stores the full blockchain history. Evidently, EtherClue is intended to complement, and not replace, vulnerability detectors. While vulnerability detectors may prevent some vulnerabilities to be deployed in production environments, in this case a blockchain, this does not mean that the code is bulletproof. On the contrary, new attacks which may be even unknown to the vulnerability detectors can lead to a late vulnerability discovery with the smart contract already deployed.

In this context, one may investigate a suspected successful exploit. Examples of suspicious activities may include the unexpected draining of an account's balance, an account getting stuck in a specific state or associated with an unusually high number of failed transactions. Traces from the suspect transactions are passed to vulnerability detectors, potentially presenting execution paths that would have been missed by them during pre-deployment analysis, or be manually inspected. Once vulnerabilities are confirmed, their location is provided to EtherClue to obtain further information about the incidents concerned. 

\noindent\textbf{Main contributions:} The contributions of this work are manifold. More precisely, first, we introduce a model for Ethereum smart contract execution covering the block, transaction and EVM levels, and upon which IoCs for the integer under/overflow, DoS, and reentrancy exploit classes are defined. Furthermore, we introduce EtherClue, a new open-source DFIR tool making use of these IoCs. 

We perform experiments with EtherClue in both synthesised and real-world use cases and transactions. The full dataset for the synthesized cases has been publicly released, along with attack transaction labels for the real-world ones. Our experimental results show that block-level IoCs can produce results quicker than at the EVM-level as EVM instruction traces grow larger, while the opposite is true for an increase in contract storage size. Also, block-level IoCs are contract-specific and less precise than those defined at the EVM level. Therefore, a balance of precision and speed depending on the application must be further assessed. Furthermore, we conducted some Ethereum mainnet case studies, demonstrating EtherClue's practicality for EVM-level IoCs. 

The rest of this paper is structured as follows: Section \ref{sec:background} provides an overview of Ethereum from the perspective of smart contract execution, as triggered by transactions, as can be observed at the EVM/transaction/block levels; Section \ref{sec:relatedwork} positions EtherClue alongside existing tools useful for digital forensics and incident response concerning Ethereum and blockchain technology in general; Section \ref{sec:etherclueioc} presents the proposed model of smart contract execution before proceeding with IoC definitions; Section \ref{sec:DFIRtool} highlights the main components of EtherClue's prototype implementation, and which is subsequently used for experimentation in Section \ref{sec:eval}. Section \ref{sec:conclusion} concludes this work by highlighting the implications of the results produced using our public dataset, as well as indicating the next steps required to bring EtherClue closer to a production-ready state.

\section{Background}
\label{sec:background}

Defining IoCs for smart contract attacks requires a characterization of their blockchain execution context. When viewing Ethereum as a world computer \cite{wood2014ethereum}, smart contracts become the programs that manipulate its global state and whose integrity is safeguarded by mainnet, Ethereum's live blockchain. Mainnet is kept synchronized and immutable by a combination of consensus and block hash chaining between participating nodes. They expose functions that, when called, as a result of transaction execution during block mining, manipulate this global state. Transactions, in turn, have their authenticity safeguarded by public-key cryptography. IoCs can either focus solely on this global state as a forensic source or consider the intermediate states resulting from individual transactions' execution, down to the individual EVM instructions as candidate forensic sources.

\subsection{Ethereum Transactions and Smart Contracts}

The most basic view of Ethereum is that of a sequence of accounts holding a value in cryptocurrency, characterizing Ethereum's global state. Accounts are identified by a 160-bit address linked to a private key used to authenticate all currency transfers requested on their behalf. \emph{Ether}, Ethereum's cryptocurrency, transfers between these so-called \emph{external accounts} take the form of transactions. Transactions identify the source/destination addresses, the value in Ether, along with a digital signature authenticating the request, among other fields. Smart contracts provide a way to define custom ways of moving Ether between accounts, enabling the development of sub-currencies (token), wallets, autonomous governance, and decentralized gambling/lottery applications \cite{whitepaper}, among others.

Listing \ref{lst:wallet} shows a simplified wallet smart contract, \texttt{Wallet.sol}, written in Solidity \cite{solidity}. A contract is itself an account on the blockchain, also holding a value in Ether, with the important difference that it includes code. \texttt{Wallet.sol} exposes two functions, \texttt{deposit()} (lines 5-12) and \texttt{withdraw()} (lines 9-18), both of which are callable by transactions with the contract's address as destination. The former is marked as \texttt{payable}, meaning it can receive an Ether transfer and the payment (\texttt{msg.value}) is recorded in the \texttt{money} (permanent) storage variable (line 3). The \texttt{withdraw()} function accepts an \texttt{amount} argument as part of the transaction's call data, resulting in a further internal message call (line 12), or at least if sufficient funds are available for the transaction's source address (\texttt{msg.sender}).

\begin{lstlisting}[label={lst:wallet}, language={Solidity}, caption={\texttt{Wallet.sol} - A Solidity wallet smart contract.}]
pragma solidity >=0.5.8;
contract Wallet {
	mapping(address => uint) money;
	/* Deposit ether with the contract-account. */
	function deposit() public payable {
		money[msg.sender] += msg.value; 
	} 
	/* Withdraw ether from the contract-account. */
	function withdraw(uint amount) public {
		uint b = money[msg.sender] - amount;
		if(b > 0) {
			(bool rv, ) = msg.sender.call{value: amount}("");
			if(rv)
			{
			    money[msg.sender] = b;
			}
		}
	}
}
\end{lstlisting}

\begin{lstlisting}[label={lst:bytecode}, language={Solidity}, caption={Bytecode snippet from the compiled \texttt{Wallet.sol}.}]
...SNIP...
SUB 	//money[msg.sender] - amount
SWAP1 	//uint b = money[msg.sender] - a...
POP 	//uint b = money[msg.sender] - a...
PUSH 0	//0
DUP2 	//b
GT 		//b > 0
...SNIP...
GAS 	//msg.sender.call{value: amount}...
CALL 	//msg.sender.call{value: amount}...
...SNIP...


\end{lstlisting}

Prior to blockchain deployment and execution, smart contracts have to be compiled to EVM bytecode and possibly assigned an initial value in Ether via one more transaction. The EVM is a stack machine, as indicated by the typical stack-manipulating opcodes on lines 2-6 of the compiled \texttt{Wallet.sol} shown in Listing \ref{lst:bytecode}. Lines 9-10 concern the internal message call issued to  withdraw via the \texttt{CALL} opcode. Of particular interest is the \texttt{GAS} opcode. This instruction reflects Ethereum's gas system, where instruction execution and memory utilization consume gas \cite{ethervm}, paid in Ether by its caller account as defined by the \texttt{STARTGAS} transaction field. Its value effectively limits the maximum number of computational steps the transaction execution is allowed to take. 

This concept is crucial for Ethereum's anti-denial of service model, preventing accidental or malicious computational wastage \cite{whitepaper}. In turn \texttt{STARTGAS} is bounded by the \texttt{GASLIMIT} value associated with Ethereum blocks \cite{wood2014ethereum}. Prior to instruction execution, the EVM checks whether the gas required to execute the instruction added to the total gas utilization so far exceeds the maximum gas allotted to the respective transaction, triggering an out-of-gas exception when exceeded. By executing the \texttt{GAS} instruction on Line 9, the amount of available gas is pushed on the stack, serving as the first argument for the subsequent \texttt{CALL} instruction. In general, instructions related to the calling of other contracts, or sending Ether to accounts, are required to forward an amount of gas to the callee to consume. Gas prices are driven by a miners' market, with transaction creators specifying gas prices and miners prioritizing those transactions with favourable prices. A number of websites, e.g. ETH Gas Station\footnote{https://ethgasstation.info/} help transaction creators to get their gas prices right.

\subsection{Ethereum State Transitions}

\begin{figure*}[!ht]
  \centering
   \includegraphics[page=2,trim = 0mm 0mm 0mm 10mm, clip, width=\linewidth]{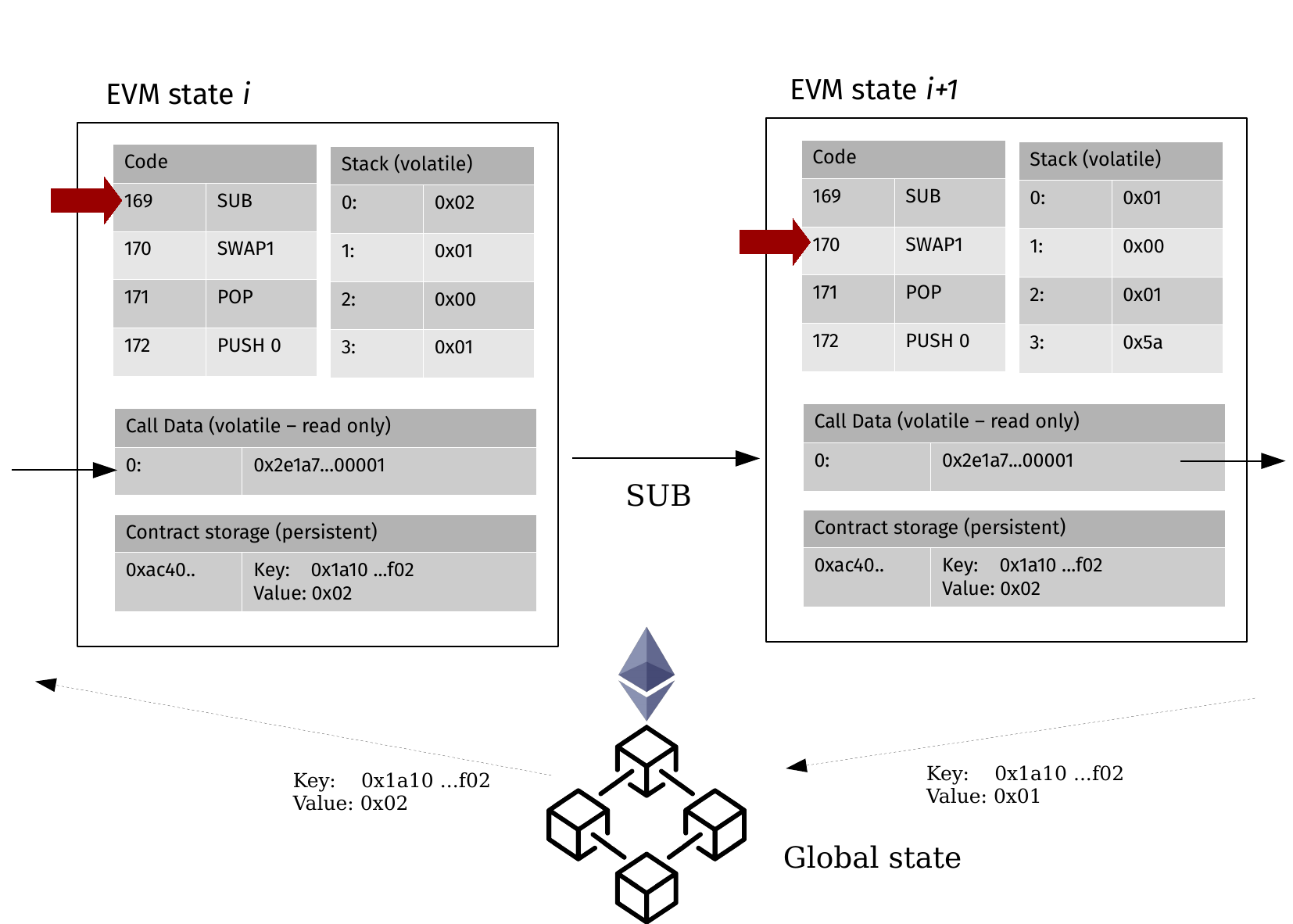}
  \caption{An EVM-level state transition triggered by instruction execution.}
  \label{fig:evmlevel}
\end{figure*}

\begin{figure*}[!ht]
  \vspace{-0.2cm}
  \centering
   \includegraphics[page=3,trim = 0mm 0mm 0mm 10mm, clip, width=\linewidth]{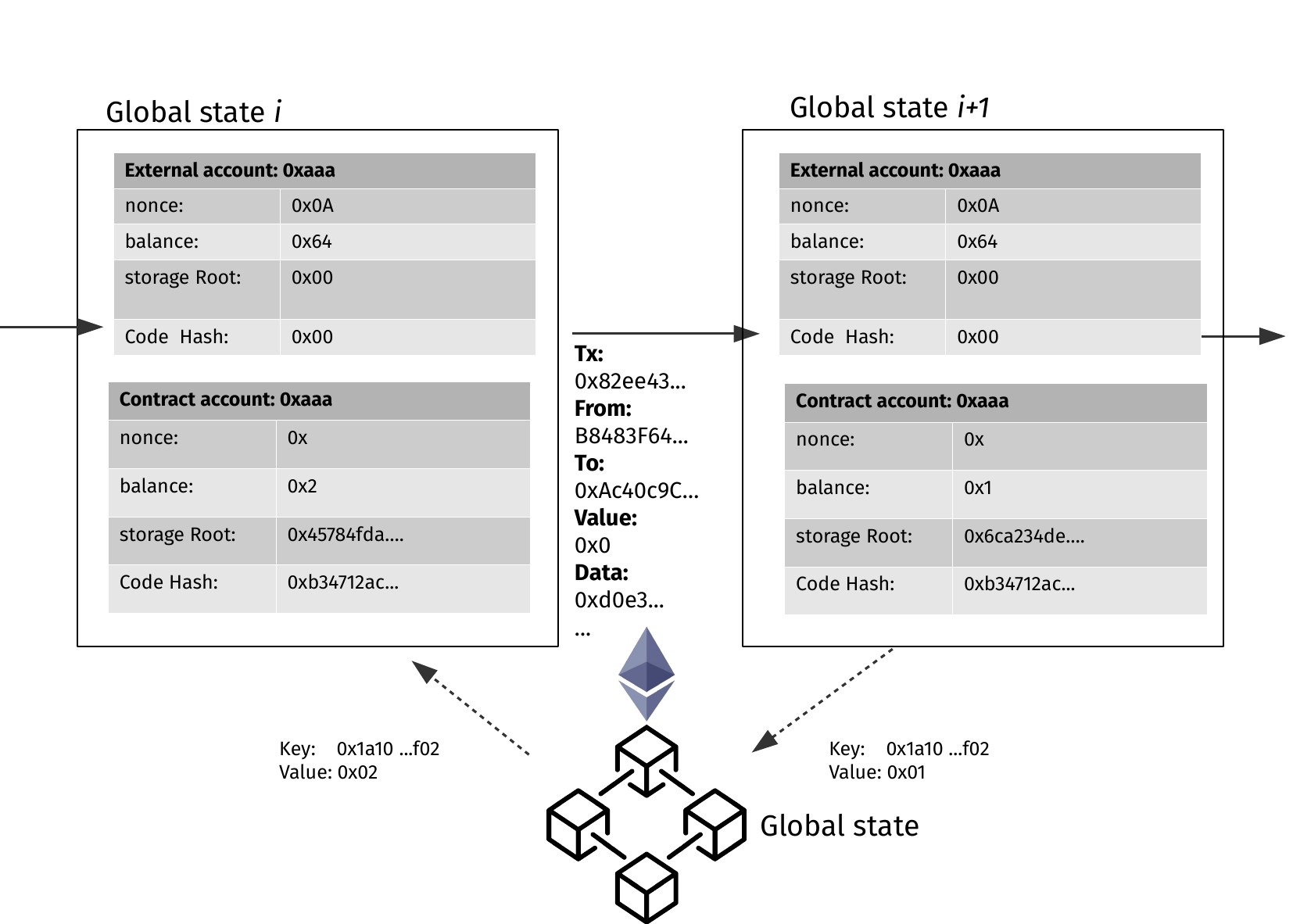}
  \caption{A transaction-level state transition triggered by transaction execution.}
  \label{fig:txlevel}
  \vspace{-0.1cm}
\end{figure*}

When considering Ethererum as a state machine, the execution of \texttt{Wallet.sol} triggers state transitions on three levels: EVM instruction, transaction, and block mining. At the EVM-level, Figure \ref{fig:evmlevel}, the contract's persistent storage and EVM's volatile memory sections characterize the execution state, while EVM instructions trigger transitions. A new state is created at the start of all external and (for some) internal calls. A read-only data-section stores the call's arguments, other transaction data, e.g. sender's address, transaction value in Ether, as well as block-related context, e.g. block's hash and timestamp. The contract storage is also made available, in our example storing \texttt{money}'s key-value pairs. Any changes to it will eventually be reflected in the global state post-successful execution. The stack area is the most dynamic, with Figure \ref{fig:txlevel} illustrating a state transition triggered by a \texttt{SUB} instruction, acting on the top of the stack during \texttt{withdraw()}'s execution.

The transaction level's state, Figure \ref{fig:txlevel}, is at a higher level of abstraction and is oblivious of the EVM level. Rather, it is concerned with transactions updating Ethereum's global state. This state is an array of externally owned and contract accounts, with the only difference with the latter being associated with contract code and their persistent storage variables. Transaction-level state transitions, comprising the initial transaction and any internal ones triggered as a consequence, manipulate accounts through transaction fields (e.g. destination account address and the cryptocurrency value fields) and contract code execution. In the case of a transaction calling \texttt{Wallet.sol}'s withdrawal function for 1 Wei (1 Ether = $10^{18}$ Wei), the post-state reflects an updated value in the \texttt{money} map. 

\begin{figure*}[!ht]
  \vspace{-0.2cm}
  \centering
   \includegraphics[page=4,trim = 0mm 2mm 0mm 30mm, clip, width=\linewidth]{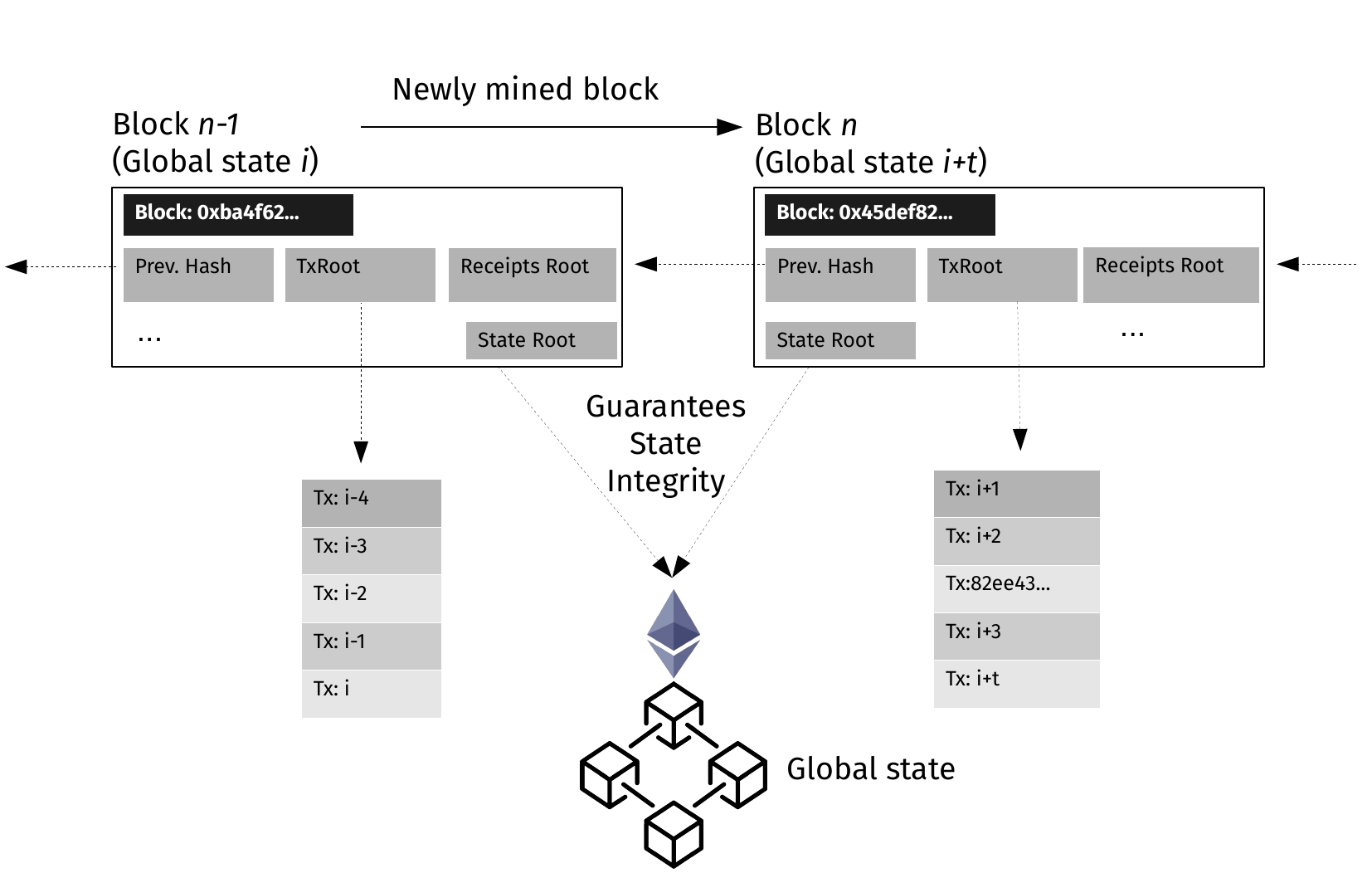}
  \caption{A block-level state transition triggered by block mining.}
  \label{fig:blklevel}
  \vspace{-0.1cm}
\end{figure*}

Transactions are executed in blocks, and they are stored as an immutable blockchain to safeguard the global state's integrity. At the block level, Figure \ref{fig:blklevel}, execution states are defined by these newly accepted blocks of transactions (Tx) and associated block meta-data, e.g. the block hash and the hash of the previous block chained to it. This level does not keep account of intermediate states generated by individual transactions or EVM-level state transitions, except for values inside transaction receipts (Rx), or log entries made by contract code. Nevertheless, through transaction tracing, typically offered by Ethereum nodes \cite{evmtracing}, it is possible to replay transaction execution up to EVM instruction level. This is achieved by recomputing (or caching) the global state at the transaction call point through blockchain traversal and accessing the contract bytecode stored as part of the global state. The global state's integrity is safeguarded by a hash, called the state root, stored as part of every block.

\subsection{Notation and Execution Semantics}
\label{sec:notation}

The semantics underpinning Oyente's symbolic execution engine \cite{luu2016making}, were chosen  among other candidates \cite{grishchenko2018semantic,hernandez2020analysis} since they provide a more convenient starting point for the required model of smart contract execution. Thus, their coverage of all three levels of smart contract execution simplifies model development.. They build upon the formalism presented in the Ethereum Yellow Paper \cite{wood2014ethereum}; referred to from now on as EYP, where the \emph{global state} made up of a sequence of all external/contract accounts is denoted as $\sigma$. In contrast, the state of a single account with address $\gamma$ is denoted as $\sigma[\gamma]$. A single external/contract account state $\sigma[\gamma]$, is the tuple $\langle balance, nonce, storageHash, codeHash \rangle$. The first two tuple elements correspond to the current value in currency owned by the account and countermeasure for replay attacks on digitally signed transactions. The last two are only relevant for contract accounts. They correspond to the root hashes of a (modified) Merkle trie structure (a key/value store with integrity protection) \cite{wood2014ethereum} storing all contract persistent variables and the hash of the smart contract code, respectively. 

Accounts states are accessible via the \emph{world state} lookup function, $\gamma \rightarrow \sigma[\gamma]$, in turn requiring access to the blockchain $BC$ to verify account state integrity. The three levels of execution, at the EVM/transaction/block levels are defined by the semantics of small-step evaluation ($\rightsquigarrow$), successful transaction execution ($\stackrel{T}{\longrightarrow}$) and big-step evaluation ($\Downarrow$), updating the EVM state ($\mu$), $\sigma$ and $BC$ respectively. Operational semantics rules are used, each describing execution at the EVM, transaction or block levels of abstraction, and structured as follows:

\vspace{3mm}
\inference[\hspace*{8em}Execution:]
{
expression_{1}(\mu|\sigma|BC,\mu'|\sigma'|BC')\\
expression_{2}(\mu|\sigma|BC,\mu'|\sigma'|BC')\\
...\\
expression_{n}(\mu|\sigma|BC,\mu'|\sigma'|BC')\\
}
{\mu|\sigma|BC \;\;\;\;\;\; \rightsquigarrow|\stackrel{T}{\longrightarrow}|\Downarrow \;\;\;\;\;\; \mu'|\sigma'|BC'}
\vspace{3mm}
\noindent where expressions describe how post-states are evaluated from pre-sates and define any constraints on the transition. At the EVM level, the semantics of a subset of the EVM, called \texttt{EtherLite}, is provided. Combined, these semantics inform the implementation of a symbolic execution engine. In turn, Oyente benefits from a symbolic execution-based approach to inspecting contracts on a path-by-path basis. For each path, a number of instruction sequence-centric conditions are applied to detect vulnerabilities. This approach is superior compared to dynamic testing, eliminating the need for input generation to maximise code coverage. However, accurate emulation of runtime execution becomes critical, warranting the need for precisely defined execution semantics. In contrast, since EtherClue operates on traces derived from concretely replayed transactions, this is no longer a requirement.

\section{Related work} 
\label{sec:relatedwork}
Post-factum detection and recovery from smart contract attacks is an area that has received much less attention as compared to proactive defences. While this is understandable, this problem cannot be ignored, especially when one considers the limitations of vulnerability detectors and the need to identify malicious or subverted accounts. DappGuard \cite{cook2017dappguard} is the earliest effort we found in this direction, with a focus on studying whether transactions and transaction receipt information extracted by blockchain exploration can help to uncover attacks. Its IoCs are referred to as attack fingerprints and compare to EtherClue's block-level IoCs. Similar approaches attempt to identify the discriminating features of accounts and associated transactions using machine learning \cite{farrugia2020detection,kumar2020detecting}. 

TxSpector \cite{zhang2020txspector}, on the other hand, is the most recent related work and, in fact, carried out in parallel to EtherClue's study. It adopts a static analysis method to work on dynamic EVM bytecode traces instead and therefore compares to our EVM-level IoCs. Instruction traces enriched with data and control flow dependencies are derived using a modified Ethereum node. Manually written logic rules are used to identify those traces, at this point represented as logical relations, generated from known attacks. TxSpector can be seen to be a generalization of Sereum \cite{rodler2018sereum}, which take a dynamic taint analysis approach to detect reentrancy exploit transactions. While Sereum was originally positioned as a real-time detector, implemented by customizing an Ethereum node, its underpinning technique could still be employed in a post-factum manner to flag transactions in a similar fashion to TxSpector. Yet its scope is restricted to just reentrancy attacks and does not offer support for writing additional detector rules or IoCs, as per TxSpector or EtherClue, respectively.

Ultimately the goals of EtherClue and TxSpector are similar, however with one crucial difference. TxSpector, while capable of detecting exploit transactions, is actually conceived as a dynamic vulnerability detector, operating on instruction traces of already-executed transactions, as opposed to considering all possible execution paths that would have to be considered by a static analysis tool. On the other hand, EtherClue is positioned to complement vulnerability detectors, following up vulnerabilities detected within deployed contracts with an investigation concerning finding out whether the contract has yet been attacked along which accounts are implicated. The result is that of much simpler detection rules focusing on execution side-effects rather than program analysis-based ones. Furthermore, EtherClue's operational semantics foundation, which already underpins Oyente's \cite{luu2016making} symbolic execution engine, can even provide a basis for automated IoC generation.

Other works have the potential to complement EtherClue during incident response. Firstly, the identified exploit transactions can be used to detect other attack transactions related to the same incident, while not necessarily being exploit transactions per se, e.g. transactions related to the setting up of malicious smart contracts or ones that transfer stolen currency to financial service contracts. DEFIER \cite{suevil} is the only such tool we are aware of, whereby given a seed set of attack transactions, it performs data analysis using code/graph similarity, as well as time series analysis based on manually inspected attack kill-chains, to identify further incident-related transactions. Two other studies \cite{bragagnolo2019towards,wu2020ethscope} address the problem of the ever-increasing size of archive nodes and the implication to blockchain analysis. Some of the explored aspects include the application of parallel map/reduce and the practice of incomplete state storage that does not compromise high-fidelity transaction replay. 

Once post-factum attack transactions are identified, the subsequent step is graceful recovery. One approach is to prepare for recovery during the smart contract programming stage using security patterns. The \emph{Emergency Stop} pattern \cite{wohrer2018smart} involves deactivating critical contract functions when certain conditions are met, leaving only the possibility to withdraw funds. The \emph{Virtual Upgrade} pattern \cite{virtupgrade} is less drastic, where critical functions are only exposed through a proxy contract that can dynamically route transactions to back-end contracts. This pattern provides flexibility, re-routing transactions to patched contracts at runtime. Automated patching can speed up this time-critical operation, reducing the opportunity window for attacks further still \cite{rodler2020evmpatch}.

Other studies have a completely different scope than EtherClue's. Using honeypots for threat analysis, for example, can shed light on little-known attack techniques \cite{cheng2019towards}. ContractLarva is a Runtime Verification tool that can prevent exploits from succeeding by instrumenting smart contracts with security property verification \cite{colombo2018contracts}. While this approach raises smart contracts' security level, it can only do so with respect to the defined properties and comes at the cost of additional gas consumption. Differently from incident response, the investigation of cryptocurrency-driven blockchains also concerns transactions related to fraud, extortion, money laundering and tax evasion during forensic accounting exercises \cite{weber2019anti,zarpala2020blockchain}. Finally, permissioned/private blockchains can be useful for chain-of-custody management of digital evidence \cite{lone2019forensic,ahmad2020blockchain}.

\section{EtherClue IoCs}
\label{sec:etherclueioc}

The model of smart contract execution encompasses all three levels of abstraction for state transitions, i.e. the EVM, transaction and block levels. The base state transitions build upon the operational semantics used for Oyente \cite{luu2016making}, while EtherClue's IoCs are defined in terms of additional constraints over the base definitions.

\subsection{Model of smart contract execution}

Oyente's \cite{luu2016making} original semantics are primarily intended for accurate emulation, providing a suitable starting point for our model. Some modifications/additions are necessary to fit EtherClue's digital forensics needs and are as follows:

\begin{itemize}
    \item Introduce a new \emph{world state} map that supports the retrieval of past global states.
    \item Provide custom definitions for blocks, transactions, and related structures, based on those found in the EYP so as to focus solely on the forensic sources of interest.
    \item Likewise, abstract away details of the block mining process.
    \item Enhance the model with transaction receipts since these constitute evidence concerning individual transaction execution.
    \item Enhance EVM-level operational semantics with a provision to indicate the location of vulnerable code.
    \item EVM-level semantics are defined over grouped instruction classes, rather than for \texttt{EtherLite} (the sub-set of EVM instructions considered by Oyente). In this manner, a single IoC can be conveniently defined over multiple EVM instructions.
    \item Provide IoC templates for the EVM and block levels to detect exploit transaction during transaction replay.
\end{itemize}

We now enlist all definitions, the execution model underpinning EtherClue's IoCs, highlighting the parts specific to EtherClue explicitly by enclosing them into a rounded-boxed border.

The \emph{global state} $\sigma$ is an array of accounts, identifiable by the 160-bit contract address $\gamma$. We also adopt the $\sigma[\gamma]_{s}$ and $\sigma[\gamma]_{c}$ notation, to conveniently refer to the actual storage and code content respectively, rather than just the root hashes, and through which to access their internal elements, e.g. $\sigma[\gamma]_{s}[varname]$.


According to the EYP:
\begin{definition} An Ethereum account  $\sigma[\gamma]$ is defined as the tuple: \[\sigma[\gamma]\stackrel{def}{=} \langle nonce,balance,storageHash,codeHash \rangle,\] where
$nonce$ denotes a fresh number to use as countermeasure against transaction replay attacks;
$balance$ denotes the account balance in Wei ($=10^{-18}$ Ether); $storagehash$ denotes the root hash for the contract storage's Merkle trie, and $codeHash$ denotes the hash for associated smart contract EVM code.
\label{def-sigma}
\end{definition}

\paragraph{EVM level} Definitions at this level are expressed in the spirit of all smarts contracts executing in the context of the current global state $\sigma$ and observable through a trace of EVM opcodes, as  derived through transactions replayed by an Ethereum node that supports the \texttt{debug\_traceTransaction}\footnote{https://geth.ethereum.org/docs/rpc/ns-debug\#debug\_tracetransaction} API call. See Table \ref{tbl:evmtransitions} for an example. An \textit{Activation Record Stack} $A$ of \textit{Activation Records} $AR$ is used to implement the mechanism of internal transaction execution. $AR$s keep track of the EVM state $\mu$ to be resumed once the current nested transaction returns control to its caller. The $A$ for an external transaction always starts empty ($\epsilon$). Whenever an exception $EXC$ is thrown, it is placed on the top of the stack, replacing the topmost $AR$, effectively returning control to its caller, that either handles $EXC$ or bubbles it up further in the call stack. An $AR$ reflects the $\mu$ that is to be reinstated following caller resumption, and therefore has to keep track of the executing code's address (only addition to Oyente's definition), the code itself and the next instruction to execute on resumption, along with stack and  volatile random-access memory's transient content, that would have been cleared out prior to internal transaction execution. Therefore $\mu$ comprises the current $A$ and the current $\sigma$ in which it executes. State transitions on this level are described by small-step evaluations ($\rightsquigarrow$), and reflect the $\mu$s obtained for traced transactions. The full definition list for the EVM-level is the following:

\begin{tcolorbox}[standard jigsaw,opacityback=0]
\begin{definition}We define an Activation Record (AR) as the tuple \[AR \stackrel{def}{=} \langle id,M,pc,l,s \rangle,\] where $id$ denotes the $\gamma$ of executing code, $M$ denotes the smart contract's EVM code, $pc$ denotes the program counter representing an offset into $M$, and $l,s$ denote the transient data in random-access volatile memory and stack memory data accordingly.
\label{def-actrec}
\end{definition}
\end{tcolorbox}

\begin{definition}Activation Records $AR$ are stored in Activation Records Stacks $A$ defined as:
\[A \stackrel{def}{=} A^{normal}| EXC \cdot A^{normal}\]
\[A^{normal} \stackrel{def}{=} AR \cdot A^{normal}| \epsilon\]
where `$\cdot$' separates the top element, an $AR$ or a thrown exception $EXC$, from the rest of the stack; $A^{normal}$ denotes an activation record for `normally halting' calls, while $\epsilon$ denotes the empty stack.
\label{def-AR}
\end{definition}

\begin{definition} For an Activation Records stack $A$ and an Ethereum account $\sigma$ providing the execution context (see Definition \ref{def-sigma}), we define an EVM state $\mu$ as the tuple: \[\mu \stackrel{def}{=} \langle A,\sigma \rangle\]

\end{definition}

\begin{definition} We abstract the transition from  an EVM state $\mu$ to an EVM state $\mu'$, as triggered by EVM opcodes inside replayed transaction traces, and call it a \textit{Small-step Evaluation} representing it as $\mu \leadsto \mu'$.  
\end{definition}

Note that in Table \ref{tbl:evmtransitions}, $a[i\mapsto v]$ evaluates to a new array identical to $a$, but storing the new value $v$ at position $i$.

\begin{table*}[!th]
\centering
\footnotesize
\caption{Example EVM-level transitions are described in terms of how specific EVM opcodes inside replayed transaction traces can be observed to alter the abstract EVM state $\mu$.}
\resizebox{\textwidth}{!}{%
\begin{tabular}{|l|l|l|l|}
\hline
  $M[pc]$ &Conditions & $\mu$ & $\mu'$  \\
  \hline 
  PUSH8 1 &-&$\langle\langle id,M,pc,l,s \rangle,\cdot A, \sigma \rangle$&$\langle\langle id,M,pc+1,l,1 \cdot s \rangle\cdot A, \sigma \rangle$ \\
  \hline
  PUSH8 5  &-&  $\langle\langle id,M,pc,l,1 \cdot s \rangle\cdot A, \sigma \rangle$ &   $\langle\langle id,M,pc+1,l,5 \cdot 1 \cdot s \rangle\cdot A, \sigma \rangle$ \\
  \hline
   &$k=5 \leftarrow pop(s)$&& \\
  SSTORE&$v=1 \leftarrow pop(s)$&  $\langle\langle id,M,pc+1,l,5 \cdot 1 \cdot s \rangle\cdot A, \sigma \rangle$ & $\langle\langle id,M,pc,l,s \rangle\cdot A, \sigma' \rangle$ \\
  &$\sigma' \leftarrow$$\sigma[id][storage][k \rightarrow v]$ &&\\
  \hline
\end{tabular}
}
\label{tbl:evmtransitions}
\end{table*}

In EtherClue, we opted to define IoCs over grouped instruction classes and require the vulnerability's location(s) and its associated parameters. This is both reasonable and practical since EtherClue's scope complements and does \emph{not} replace vulnerability detectors. Furthermore, IoC definitions can focus solely on the identification of exploit side-effects, without being concerned with defining the nature of the vulnerability per se. The notation we make use of, $\otimes^{NAME}_{PARAM_{1...N}}$, identifies EVM instruction classes by $NAME$. The parameter list ${PARAM_{1...N}}$ provides a vulnerability-specific context for the instruction class in question. For example, $\otimes^{INT}$ groups multiple integer arithmetic instructions. In terms of Solidity source language constructs, this grouping ranges over all possible integer sub-types (\texttt{UINT8-256}, \texttt{INT8-256}). Therefore a parameter list is required to define the lower/upper bounds for specific sub-types. These parameters assume the availability of the (possibly decompiled) source code. On the other hand, $VULN\_LOCS$, a set of code offsets belonging to a detected vulnerability, is instantiated with the code offsets from $M$ associated with it, and where every opcode belongs to $\otimes^{NAME}$.
 
To provide a common template, EVM-level IoCs are defined as a sequence of constraint expressions, defined over $\mu$, $\mu'$, and $VULN\_LOCS,PARAM_{1...N}$, qualifying those side-effects that render a small-step evaluation to be considered an exploit step. When all predicate expressions evaluate to true for a particular small-step evaluation, the corresponding transaction is flagged as an exploit during replayed execution. More formally:

\begin{tcolorbox}[standard jigsaw,opacityback=0]
\begin{definition}We define an \emph{EVM-level IoC template} as an EVM instruction class:\\
{
\inference[\hspace*{6em}$\otimes^{NAME}_{PARAM_{1...N}}$:]
{
VULN\_LOCS \leftarrow M \text{ offsets}\\
expression_{1}(\mu,\mu',VULN\_LOCS,PARAM_{1...N})\\
expression_{2}(\mu,\mu',VULN\_LOCS,PARAM_{1...N})\\
...\\
expression_{n}(\mu,\mu',VULN\_LOCS,PARAM_{1...N})\\
}
{\mu \rightsquigarrow \mu'}
}
\\\\where $M \text{ offsets}$ is a set of code offsets belonging to a vulnerability; $VULN\_LOCS$ is a symbolic constant representing the above set of code offsets; $\mu$,$\mu'$ denote EVM-level pre/post-states; $\otimes^{NAME}$ is an EVM instruction class that matches the IoC; $PARAM_{1...N}$ represent EVM instruction class parameters; 
$expression_{i}$ are operational semantics expressions defined over $\mu,\mu',VULN\_LOCS,PARAM_{1...N}$, with an IoC match occurring whenever all predicate expressions evaluate to true.

\label{def:evmtemplate}
\end{definition}
\end{tcolorbox}

\paragraph{Transaction level} At this level, we consider the $sender$, $to$, $value$ and $data$ elements of successful transactions $T$ as evidence sources. These elements correspond to the source/destination addresses, amount of cryptocurrency being transferred, and a data field identifying the contract's public function and arguments. Transactions are executed as a sequence of $0$ (when a destination is also an external account) or more small-step evaluations $\leadsto^{*}$. State transitions $\stackrel{T}{\longrightarrow}$ at this level operate on $\sigma$, transferring cryptocurrency corresponding to $T$'s $value$ to $\sigma[to]$'s $balance$. Further updates to $\sigma$ may occur as a result of executing $\sigma[to]_{c}$ on termination of a small-step evaluation $\leadsto^{*}$. The finalized definitions comprise a sub-set of transaction attributes defined in the EYP and the definitions provided by Oyente for transaction execution, either ending successfully or with an exception. In the latter case, the currency associated with the transaction is not transferred, and the changes to the global state are not committed. Definition \ref{def-transaction-successful}, while maintaining the original semantics of transaction execution, its defined to reflect modifications in its primitive definitions.

\begin{tcolorbox}[standard jigsaw,opacityback=0]
\begin{definition} We define a Transaction $T$ in terms of a subset of the EYP's definition, as the tuple $T \stackrel{def}{=} \langle sender, to, value, data \rangle$, where $sender$ denotes sender's account address $\gamma$, 
$to$ denotes the target's account address $\gamma'$, 
$value$ denotes the transaction amount in Wei, and $data$ denotes the call data. That is an array comprising the hash of a public function exposed by the target $to$ followed by the function arguments
\label{def-transaction}
\end{definition}
\end{tcolorbox}

\begin{definition}Transaction Execution $\stackrel{T}{\longrightarrow}$:\\\\
\inference[\hspace*{6em}Tx-{\tiny{SUCCESS}}:]
{
\langle sender,to,value,data \rangle \leftarrow T \\
M \leftarrow \sigma[to]_{c} \\
\sigma' \leftarrow \sigma[to][balance \mapsto (\sigma[to][balance] + value)] \\
\langle\langle to,M,0,data,\epsilon \rangle\cdot \epsilon, \sigma' \rangle\leadsto^{*}\langle \epsilon,\sigma'' \rangle 
}
{\sigma \stackrel{T}{\longrightarrow}  \sigma''}
\\\\\\\\
\inference[\hspace*{6em}Tx-{\tiny{EXCEPTION}}:]
{
\langle sender,to,value,data \rangle \leftarrow T \\
M \leftarrow \sigma[to]_{c} \\
\sigma' \leftarrow \sigma[to][balance \mapsto (\sigma[to][balance] + value)] \\
\langle\langle to,M,0,data,\epsilon \rangle\cdot \epsilon, \sigma' \rangle\leadsto^{*}\langle EXC\cdot\epsilon,\sigma'' \rangle 
}
{\sigma \stackrel{T}{\longrightarrow}  \sigma}

\;\;\\\\where Tx-{\tiny{SUCCESS}} represents a successfully completed transaction; while Tx-{\tiny{EXCEPTION}} represents a failed transaction ending with a thrown exception; $\sigma$, $\sigma''$ represent the pre/post (global) states; while $\sigma'$ represents the interim (global) state which stores the new target account's balance prior to smart contract execution; $\leadsto^{*}$ represents a sequence of 0 or more small-step evaluations ($\rightsquigarrow$), depending on whether the target address $to$ is a smart contract account or not; and $EXC$ is a thrown exception.
\label{def-transaction-successful}
\end{definition}

While the transaction-level definitions are needed by the block-level ones, we deem IoCs at this level to be unnecessary. Detecting IoCs at this level would first and foremost accessing the state $\sigma$ and $\sigma''$, upon which IoCs at this level would be defined. While transaction receipts at the block level (see Definition \ref{def-transaction-rec}) can shed some light on the make-up of these states, their complete characterization is not. Rather, the only manner with which $\sigma$ and $\sigma''$ can be computed is by executing $\leadsto^{*}$. Since this amounts to the same processing carried out at the EVM level, which also has access to $\sigma$ and $\sigma''$, EVM-level detection trumps the transaction level at no additional cost.

\paragraph{Block level} This level is concerned with the execution of batched transaction sequences $TXs$, as chosen out of a transaction pool set $\Gamma$, during successful block mining. Successful transaction execution results in transaction receipts $Rx$ (we only consider the log entries $logs$ and transaction status codes $statusCode$) to be added to the block's receipt sequence $RXs$. A newly mined block $\beta$, therefore, performs a big-step evaluation, transitioning from the global state $\sigma_{0}$ directly to $\sigma_{n}$ after having batch-executed $n$ transactions. Finally, $\beta$ is appended to the blockchain $BC$, comprising a sequence of all the blocks mined so far. Overall we ignore details of the consensus protocol considered to not contribute as a forensic evidence source,  e.g. details of the block mining process, such as proof-of-work, or the propagation of successfully mined blocks across the entire Ethereum peer network. However, we are not concerned with executing the next block of transactions here, but rather with being able to fetch any past $\sigma_{0}$/$\sigma_{n}$ pair and verify whether the corresponding $\Downarrow$ constitutes a transition during which an exploit transaction occurred. To add this capability to our model, we enhance the notion of the World State $\omega$ mapping with the ability to take any state root (a hash of the global state corresponding to a block $\beta$) from the past and return its corresponding $\sigma_{stateRoot}$. This enhanced definition naturally also applies to EVM-level IoCs, and through which they obtain the ability to replay transaction traces in the first place. Therefore we say that EtherClue operates on an Archive Node $\langle BC,\omega\rangle$. The resulting definitions, omitting details unused by EtherClue and adding the concept of an archive node, are as follows.

First, we use Oyente's definition for a Transaction sequence:
\begin{definition}A Transaction sequence $TXs$ is an ordered set of $n$ transactions \[TXs\stackrel{def}{=} (T_1, \ldots, T_n)\]
\end{definition}

\begin{tcolorbox}[standard jigsaw,opacityback=0]
\begin{definition}We define a Transaction Pool $\Gamma$ as a sequence of transactions $T_i$ of arbitrary size $m$: \[\Gamma \stackrel{def}{=} \{T_0, \ldots, T_m\}\] 

from which $TXs$ of size $n$ can be chosen in any order and $m > n$.

\end{definition}

\begin{definition}We define a Transaction Receipt $R$ as a tuple: \[R \stackrel{def}{=} \langle logs, statusCode \rangle,\] where $logs$ denote the event log entries as created by the EVM \texttt{LOGX} opcodes and $statusCode$ denotes the transaction status code (success or failure).
\label{def-transaction-rec}
\end{definition}

\begin{definition}We define a Receipt Sequence $RXs$ as an ordered sequence of $n$ Transaction Receipts: \[RXs \stackrel{def}{=} (R_1, \ldots, R_n)\]
\end{definition}
\end{tcolorbox}

We use a subset of the Block definition from EYP as it is deemed relevant as a forensic source.
\begin{tcolorbox}[standard jigsaw,opacityback=0]
\begin{definition}We define a Block $\beta$ as the tuple: \[\beta \stackrel{def}{=} \langle parent, TXs, RXs, stateRoot \rangle,\] where $parent$ refers to the block's parent block through its hash, 
$stateRoot$ denotes the root hash of the Merkle trie storing $\sigma$, and $TXs$, $RXs$ denote a Transaction and Receipt Sequence, respectively.
\label{def-block}
\end{definition}  

\begin{definition} We define a Blockchain $BC$ as a sequence of $n$ blocks: \[BC\stackrel{def}{=} (\beta_0, ..., \beta_n)\]
\end{definition}

\begin{definition}We define a World State $\omega$ as: \[\omega \stackrel{def}{=} stateRoot \mapsto \sigma_{stateRoot},\] where
$\sigma_{stateRoot}$: $\sigma$ for some block $\beta$ with $\beta[stateRoot]=stateRoot$
\end{definition}

\begin{definition} We define an Archive Node $\mathcal{A}$ as a tuple of a blockchain $BC$ and a world state $\omega$: \[\mathcal{A}=\langle BC, \omega \rangle\]
\end{definition}
\end{tcolorbox}

\begin{tcolorbox}[standard jigsaw,opacityback=0]
\begin{definition} For a given Transaction Pool $\Gamma$ from which a transaction sequence $TXs$ is selected by a miner at its discretion, and using $\epsilon$ to denote and empty sequence and $...$ to delimit the first and last elements of a sequence; we define a \textit{Big-step evaluation}, denoted as $\Downarrow$ in the context of blockchain $BC$, and generating the Receipt Sequence $RX$, as :\\\\
\inference[\hspace*{6em}BIG-STEP:]
{
TXs \leftarrow select(\Gamma)\\
(T_1, ..., T_n) \leftarrow TXs \\
Rx\leftarrow \epsilon\\
\beta_{parent} \leftarrow last(BC) \\
\sigma_{0} \leftarrow \omega[\beta_{parent} [stateRoot]] \\
\forall i, \quad 1 \leq i \leq n: \{ \\ \sigma_\text{i-1} \xrightarrow{T_i}  \sigma_i, Rx[i] \leftarrow generateReceipt(T_{i})\}\\
\beta \leftarrow \langle \beta_{parent}, TXs, RXs, rootHash(\sigma_{n}) \rangle \\
\omega' \leftarrow \omega[\beta[stateRoot] \mapsto \sigma_n] \\
}
{
\langle BC,\sigma_{0}, \omega \rangle \Downarrow \langle BC \cdot \beta, \sigma_{n}, \omega' \rangle 
}
\\\\where $\sigma_{0},\sigma_{n}$ denote  the global state prior/post $\beta$, respectively; $BC \cdot \beta$ denotes the appending of block $\beta$ to $BC$; 
$\omega,\omega'$ denote the world map updated to reflect $\beta$'s state root; $select()$ denotes the miner's process of choosing transactions from $\Gamma$ at its discretion; and 
$last()$ denotes the function that retrieves the last block ($\beta$) from blockchain $BC$; $generateReceipt()$ returns a sequence of event logs for some $T$; while $rootHash()$ denotes the function that computes the state root for some global state $\sigma$.
\label{def-bigstep}
\end{definition}    
\end{tcolorbox}

Defining block-level IoCs directly on the notion of a \emph{Big-step evaluation}, as per Definition 3.18, is however counterproductive since it is requires replaying all the transactions belonging to the block $\beta$ under investigation. In effect this approach would render block-level analysis completely useless since it requires redoing all the trace generation work carried out at the EVM level. Rather, beginning from a block $\beta$ we want to investigate, we make use of $\omega$ computed during a big-step evaluation to lookup both the required $\langle BC,\sigma_{0}\rangle$ and its corresponding $\langle BC\cdot \beta,\sigma_{n}\rangle$, as opposed to deriving the latter through full transaction replay. The net effect is that of coarse-grained replay, where the side-effects at the blockchain and the $BC$ and the global state ($\sigma$) are made available at block batches i.e. for $\langle BC,\sigma_{0}\rangle$ and $\langle BC \cdot \beta, \sigma_{n} \rangle$ but not the interim $\sigma_{i}$, or any $\mu$ for that matter. To define the required partial replay, we make use of the Big-step lookup $\Updownarrow$: 

\begin{tcolorbox}[standard jigsaw,opacityback=0]
\begin{definition} For a given block $\beta$ in blockchain $BC$ containing $n$ transactions, we define its associated Big-step lookup $\Updownarrow$ as:\\\\
\inference[\hspace*{8em}BIG-STEP-LOOKUP:]
{
\beta_{parent} \leftarrow \beta[parent] \\
\sigma_{0} \leftarrow \omega[\beta_{parent} [stateRoot]] \\
\sigma_{n} \leftarrow \omega[\beta [stateRoot]] \\
}
{
\langle BC,\sigma_{0}\rangle \Updownarrow \langle BC \cdot \beta, \sigma_{n} \rangle 
}
\\\\where $\beta_{parent}$ represents $\beta$'s parent as retrieved through its block hash stored in  $\beta[parent]$; $\sigma_{0}$ and $\sigma_{n}$ are the global states corresponding to $\beta_{parent}$ and $\beta$ respectively, with $\sigma_{n}$ reflecting a global state update once $\beta$'s $n$ transactions get executed; both $\sigma_{0}$ and $\sigma_{n}$ are retrieved, rather than computed, through the world map $\omega$.
\label{def-bigsteplookup}
\end{definition}    
\end{tcolorbox}

As for a common template, block-level IoCs are made up of expressions defined over historical $\sigma_{0}$/$\sigma_{n}$ pairs, in turn, derived using the most recent $\omega$ and their corresponding $\beta$s. A block is flagged as containing an exploit transaction when all predicate expressions evaluate to true.

\begin{tcolorbox}[standard jigsaw,opacityback=0]
\begin{definition} Block-level IoC template:\\\\
\inference[\hspace*{8em}Vulnerability:]
{
expression_{1}(\sigma_{0}, \sigma_{n},\beta_{parent},\beta)\\
expression_{2}(\sigma_{0}, \sigma_{n},\beta_{parent},\beta)\\
...\\
expression_{n}(\sigma_{0}, \sigma_{n},\beta_{parent},\beta)\\
}
{
\langle BC,\sigma_{0}, \rangle \Updownarrow \langle BC \cdot \beta, \sigma_{n}, \rangle 
}
\\\\where $Vulnerability$ is a vulnerability class name to which the IoC belongs to; $\beta$,$\beta_{parent}$ denote the block under investigation and its parent, respectively; $\sigma_{0}$,$\sigma_{n}$ are the pre/post global states respectively; $\langle BC,\sigma_{0}, \rangle$ denotes the block-level pre-state comprising $BC$ prior to $\beta$'s addition and its corresponding global state $\sigma_{0}$; $\langle BC \cdot \beta, \sigma_{n}, \rangle$ denotes the block-level post-state reflecting the new global state $\sigma_{n}$ resulting from when $beta$ is appended to $BC$ and its $n$ transaction executed; $expression_{i}$ are operational semantics expressions defined over $\sigma_{0}$, $\sigma_{n}$, $\beta_{parent}$, $\beta$, with an IoC match occurring whenever all predicate expressions evaluate to true.
\label{def:blocktemplate}
\end{definition}
\end{tcolorbox}

\subsection{IoCs for exploit transactions}

In what follows, we use some vulnerable code snippets from \cite{exploits} to  illustrate how our IoCs are derived using IoC templates from Definitions \ref{def:evmtemplate} and \ref{def:blocktemplate}.

\noindent\textbf{Arithmetic over/underflow.} Listing \ref{lst:overflow} is a code snippet with integer under/overflow vulnerabilities on lines 8 and 9, respectively, as a consequence of the absence of bounds checking in both cases. 

\begin{lstlisting}[label={lst:overflow}, language={Solidity}, caption={Integer over/underflow in \texttt{Exchange.sol}.}]
mapping (address => uint256) public balanceOf;

// INSECURE
function transfer(address _to, uint256 _value) {
    /* Check if sender has balance */
    require(balanceOf[msg.sender] >= _value);
    /* Add and subtract new balances */
    balanceOf[msg.sender] -= _value;
    balanceOf[_to] += _value;
}
\end{lstlisting}

The EVM-level IoC, shown in Definition \ref{def-evmoverflowioc}, is defined over instruction class grouping all integer arithmetic operation $\otimes^{INT}$. While this IoC applies to any over/underflow IoC, the $TYPE$ parameter is contract-specific, identifying the lower/upper bounds of the integer type concerned. In the case of Listing \ref{lst:overflow}, we are concerned with an overflow in the value of the \texttt{balanceOf} mapping in the contract's storage, which is of type \texttt{uint256}. Therefore $TYPE$ in this case has to be instantiated as $\langle0, 2^{256}-1\rangle$. Similarly, $VULN\_LOC$ is a place-holder for the instruction offsets related to the corresponding integer arithmetic opcodes. Any $\otimes^{INT}$ in this region is checked for an out-of-bounds condition, verifying whether $\rightsquigarrow$ results in an over/underflow. By first casting, the integer operation to its $\mathbb{Z}$ equivalent ($\odot^{INT}$), exploit verification reduces to checking whether $\odot^{INT}$'s evaluation falls outside of the $TYPE$'s range.

\begin{tcolorbox}[standard jigsaw,opacityback=0]
\begin{definition} EVM-level integer over/underflow IoC.\\\\
\inference[\hspace*{8em}$\otimes^{INT}_{TYPE}$:]
{
pc \in VULN\_LOC\\
a',b' \in \mathbb{Z} \leftarrow \text{cast}(a,b)\\
v' \leftarrow a' \odot^{INT} b' \\
v' < TYPE.min \lor v' > TYPE.max\\
v  \in TYPE \leftarrow \text{cast}(v')\\
}
{
{\begin{array}{@{}c@{}} \langle\langle id,M,pc,l, a \cdot b \cdot s \rangle\cdot A, \sigma \rangle  \\ \leadsto \langle\langle id,M,pc+1,l, v \cdot s \rangle \cdot A, \sigma \rangle  \end{array}}
}
\\\\where $\otimes^{INT}\stackrel{def}{=} \texttt{ADD | MUL | SUB | SDIV | ADDMOD | MULMOD | EXP }$;\\ $TYPE.min/max$: define the lower/upper bounds for the EVM integer type of a given contract; $\odot^{INT}$ maps $\otimes^{INT}$ to its (infinite) $\mathbb{Z}$ counterpart; and $cast()$ maps arguments between EVM integer types and $\mathbb{Z}$.
\label{def-evmoverflowioc}
\end{definition}  
\end{tcolorbox}

Example \ref{def-blockoverflowioc} describes the block-level IoC, with the out-of-bounds condition in Listing \ref{lst:overflow} detectable at the contract storage level, rendering it contract-specific. In this case, the expressions are defined over the content of block $\beta$, and the global states prior and post its mining, $\sigma_{0}$ and $\sigma_{n}$ respectively. The expressions constraining the coarsely replayed $\Downarrow$, represented by $\Updownarrow$, check for the existence of a $T$ from $\beta$'s $TXs$ which: has \texttt{Exchange.sol}'s contract as its destination; compares the pre/post balances ($bal$ and $bal'$) of the $T$'s $sender$, and that of the account specified in the call data argument $\_to$. Given that the \texttt{transfer} function is intended to shift funds from the $sender$'s to $\_to$ accounts, the side-effect of an over/underflow is one where either the prior's balance ends up increasing or vice-versa. When all these expressions evaluate to true, the block in question is flagged as containing an exploit transaction.

\begin{tcolorbox}[standard jigsaw,opacityback=0]
\begin{example} Contract-specific block-level integer over/underflow IoC.\\\\
\inference[\hspace*{4em}o/u\_flow:]
{
TXs \leftarrow \beta[TXs]\\
\exists T \in TXs: \{\\
\langle sender,to,value,data\rangle \leftarrow T\\
to == \gamma_{Exchange.sol}\\
bal \leftarrow \sigma_{0}[to]_{s}[balanceOf]\\
bal' \leftarrow \sigma_{n}[to]_{s}[balanceOf]\\
\_to \leftarrow T[data][to]\\
bal'[sender] > bal[sender] \lor bal'[\_to] < bal[\_to] \}
}
{
{\langle BC, \sigma_{0} \rangle \Updownarrow \langle BC \cdot \beta, \sigma_{n} \rangle}
}
\\\\where $\gamma_{Exchange.sol}$ denotes \texttt{Exchange.sol}'s contract account address
\label{def-blockoverflowioc}
\end{example}  
\end{tcolorbox}

\noindent\textbf{DoS with (unexpected) revert.} Listing \ref{lst:dosrvrt} is a highest-bidder type of contract, that on receipt of a bid exceeding the current highest one, the contract-held funds are transferred back to the old bidder. Line 9, however, can be abused by an attacker making use of a destination contract account with a fallback payable function that always reverts. Line 9 therefore, always fails, pulling off a DoS on \texttt{Auction.sol}.

\begin{lstlisting}[label={lst:dosrvrt}, language={Solidity}, caption={DoS vulnerability in \texttt{Auction.sol}.}]
// INSECURE
contract Auction {
    address currentLeader;
    uint highestBid;

    function bid() payable {
        require(msg.value > highestBid);

        require(currentLeader.send(highestBid)); // Refund the old leader, if it fails then revert

        currentLeader = msg.sender;
        highestBid = msg.value;
    }
}
\end{lstlisting}

In this case, the EVM-level IoC in Definition \ref{def-dos} captures the small-step  calculation ($\rightsquigarrow$) resulting in a failed \texttt{CALL} (returns 0), and which happens specifically at the vulnerable location. Temporarily, we ignore the possibility of a similar issue happening with a \texttt{DELEGATECALL}. By definition, we do not expect that a \texttt{STATICCALL} can ever result in a DoS of this type. No additional parameters apply in this case. 

\begin{tcolorbox}[standard jigsaw,opacityback=0]
\begin{definition} EVM-level DoS with (unexpected) revert IoC.\\\\
\inference[\hspace*{10em}$\otimes^{CALL}$:]
{
pc \in VULN\_LOC\\
r=0
}
{
{\begin{array}{@{}c@{}} \langle\langle id,M,pc,l, s \rangle\cdot A, \sigma \rangle  \\ \leadsto \langle\langle id,M,pc+1,l, r\cdot s \rangle \cdot A, \sigma \rangle  \end{array}}
}
\\\\where $\otimes^{CALL}\stackrel{def}{=}\texttt{CALL}$.
\label{def-dos}
\end{definition}  
\end{tcolorbox}

At the block level Example \ref{def-blockdos} captures a Big-step lookup ($\Updownarrow$) where, due to the vulnerability in Listing \ref{lst:dosrvrt}, despite comprising a transaction value higher than the current highest bid, the highest bid remains the same. Once again, we note its contract-specific nature and that block-level flagging still applies.

\begin{tcolorbox}[standard jigsaw,opacityback=0]
\begin{example} Contract-specific block-level DoS with (unexpected) revert IoC.\\\\
\inference[\hspace*{12em}DoS:]
{
TXs \leftarrow \beta[TXs]\\
\exists T \in TXs: \{\\
\langle sender,to,value,data\rangle \leftarrow T\\
to == \gamma_{Auction.sol}\\
hbid \leftarrow \sigma_{0}[to]_{s}[highestBid]\\
hbid' \leftarrow \sigma_{n}[to]_{s}[highestBid]\\
value > hbid \land hbid'==hbid \}
}
{
{\langle BC, \sigma_{0} \rangle \Updownarrow \langle BC \cdot \beta, \sigma_{n} \rangle}
}
\\\\where $\gamma_{Auction.sol}$: \texttt{Auction.sol}'s contract account address
\label{def-blockdos}
\end{example}  
\end{tcolorbox}

\noindent\textbf{Reentrancy.} Back to \texttt{Exchange.sol}, Listing \ref{lst:reentrancy} this time shows a reentrancy issue in line 6. The contract wrongly assumes that line 8 will always be called prior to the same sender account attempting a further withdrawal. Yet, this is perfectly possible by a callback to \texttt{withdrawBalance()} performed by a malicious payable fallback function in the called contract. 

\begin{lstlisting}[label={lst:reentrancy}, language={Solidity}, caption={Reentrancy vulnerability in \texttt{Exchange.sol}.}]
// INSECURE
mapping (address => uint) private userBalances;

function withdrawBalance() public {
    uint amountToWithdraw = userBalances[msg.sender];
    (bool success, ) = msg.sender.call.value(amountToWithdraw)(""); // At this point, the caller's code is executed, and can call withdrawBalance again
    require(success);
    userBalances[msg.sender] = 0;
}
\end{lstlisting}

The IoC reflects this exploit in Definition \ref{def-reentr}, where during a small-step evaluation that updates the contract storage (and therefore $\sigma$), the activation record stack $A$ includes at least one entry; besides the one top, with the same $id$ (the contract address). This IoC evaluating to true implies that when the sender's balance is being zeroed out (line 8), this step is not being performed by the outermost call but rather as a result of an internal transaction. This is exactly the scenario that the contract's programmer was wrongly assuming as not being possible.

\begin{tcolorbox}[standard jigsaw,opacityback=0]
\begin{definition} EVM-level reentrancy IoC.\\\\
\inference[\hspace*{10em}$\otimes^{STORAGE}$:]
{
pc \in VULN\_LOC\\
\exists a\in A : a[id]=id
}
{
{\begin{array}{@{}c@{}} \langle\langle id,M,pc,l, v\cdot s \rangle\cdot A, \sigma \rangle  \\ \leadsto \langle\langle id,M,pc+1,l, s \rangle\cdot A, \sigma' \rangle  \end{array}}
}
\\\\where $\otimes^{STORAGE}\stackrel{def}{=}\texttt{SSTORE}$.
\label{def-reentr}
\end{definition}  
\end{tcolorbox}

At the block level, the IoC in Example \ref{def-blockreentr} identifies the event of a withdrawal transaction resulting in a $\sigma_{n}$ where the contract's balance ($contractval'$) does not tally with the correct user balance withdrawal ($\neq contractval-user$), possibly even fully emptied, as a consequence of the vulnerability in Listing \ref{lst:reentrancy}.

\begin{tcolorbox}[standard jigsaw,opacityback=0]
\begin{example} Contract-specific block-level reentrancy IoC.\\\\
\inference[\hspace*{10em}Reentr:]
{
TXs \leftarrow \beta[TXs]\\
\exists T \in TXs: \{\\
\langle sender,to,value,data\rangle \leftarrow T\\
to == \gamma_{Exchange.sol}\\
contractval \leftarrow \sigma_{0}[to][balance]\\
contractval' \leftarrow \sigma_{n}[to][balance]\\
user \leftarrow \sigma_{0}[to]_{s}[userBalances][sender]\\
contractval' \neq (contractval - user) \}
}
{
{\langle BC, \sigma_{0} \rangle \Updownarrow \langle BC \cdot \beta, \sigma_{n} \rangle}
}
\\\\where $\gamma_{Exchange.sol}$: \texttt{Exchange.sol}'s contract account address
\label{def-blockreentr}
\end{example}  
\end{tcolorbox}

\section{The \texttt{EtherClue} tool}
\label{sec:DFIRtool}

\begin{figure*}[!ht]
  \centering
   \includegraphics[page=2,trim = 0mm 0mm 0mm 0mm, clip, width=\linewidth]{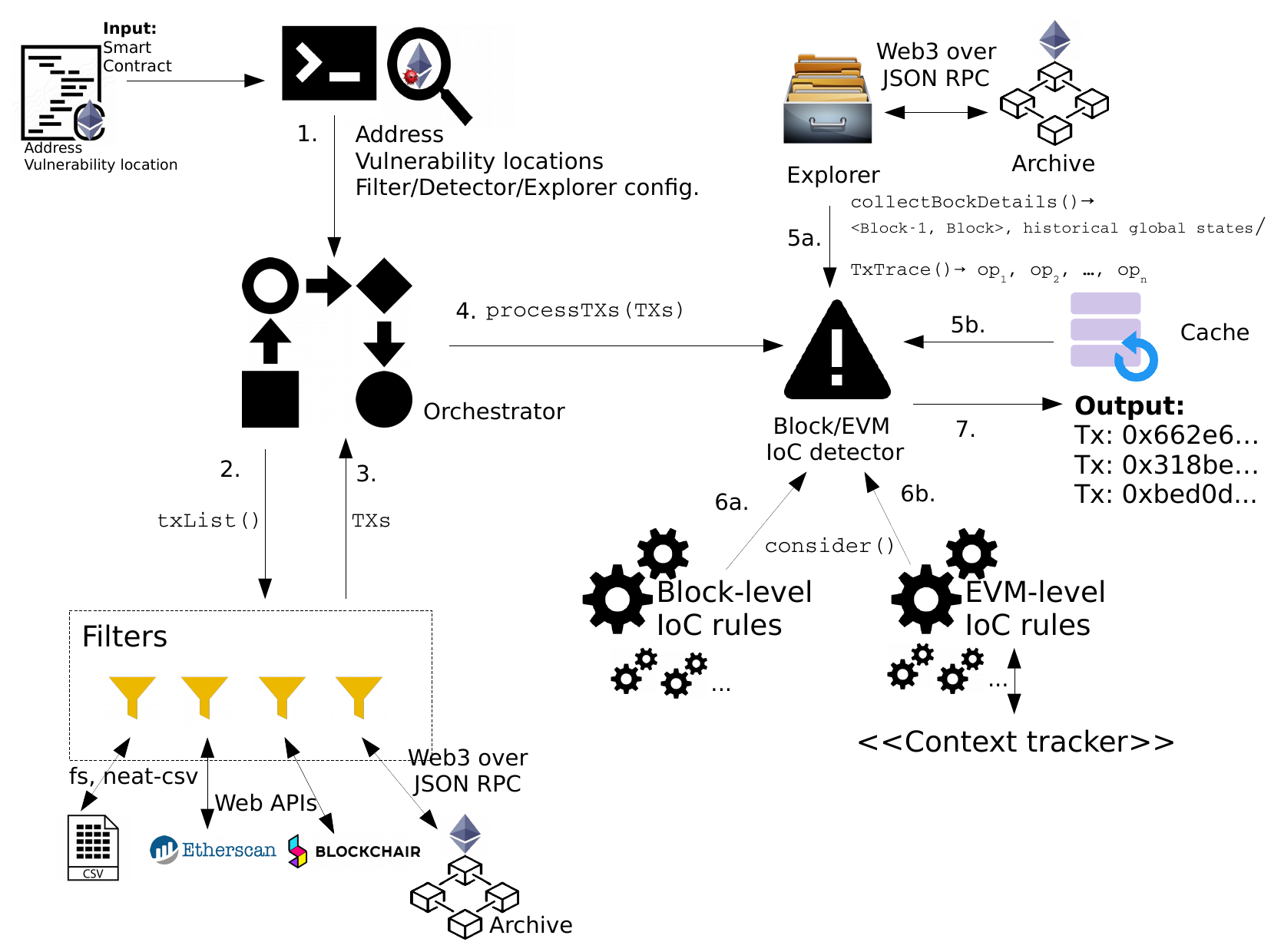}
  \caption{EtherClue's architecture and the main investigation workflow steps.}
  \label{fig:arch}
\end{figure*}

The EtherClue prototype is written in JavaScript, and its codebase is open-source\footnote{\url{https://gitlab.com/simonjaquilina/etherclue}}. It is intended to be connected either to a private Ethereum node or to a public cloud service. In either case, the nodes must be full archive nodes, meaning that they contain a full copy of mainnet, as they have to process all the code and transactions and not a fragment of them like, e.g. ArchiveNode\footnote{http://archivenode.io}. In the case of EVM-level detection, the nodes must offer full transaction tracing capabilities, i.e. single-stepping over EVM instructions with full access to the EVM state. Its main components and their corresponding interfaces are depicted in Figure \ref{fig:arch}. Once a vulnerable contract is identified post-deployment, its blockchain address and the corresponding vulnerable bytecode offsets are passed to EtherClue via its user interface. Additionally, further parameters such as filters, IoC detectors, and blockchain explorer components, are incorporated to automate the investigation accordingly. These main components, along with supporting ones, are described in the next paragraphs.

The \textbf{Orchestrator} receives user inputs and loads the necessary components, followed by coordinating the first four steps of an investigation workflow. Given the address of a vulnerable contract among the input parameters (step 1) it proceeds to call the specified \emph{filter}'s \texttt{txList()} interface to obtain the relevant transactions (\texttt{TXs}) (steps 2 and 3). All those external and internal transactions featuring the vulnerable contract's address as their target comprise the first automatic filtering stage. Additionally, any other filter parameters are also applied, e.g. the transactions' target public function. The retrieved transactions are then used to feed the block or EVM-level \emph{IoC detectors} (step 4). IoC detectors firstly obtain the required block or EVM-level information from an \emph{explorer} component, by calling its \texttt{collectBlockDetails()}, and possibly also \texttt{TxTrace()}, or by retrieving cached responses (steps 5a and 5b respectively). The obtained block information or transaction traces are sent to the \emph{IoC rule} components (steps 6a or 6b) by calling their exposed \emph{consider()} method, flagging any exploits. Finally, EtherClue outputs the list of any exploit blocks or transactions found (step 7).

\textbf{Filters.} We have implemented four filters: The first one connects to any Ethereum archive node listening over a JSON-RPC\footnote{https://geth.ethereum.org/docs/rpc/server} interface using \texttt{web3.js}\footnote{https://web3js.readthedocs.io}. Two additional filters are designed to connect to Etherscan\footnote{https://etherscan.io/apis} and Blockchair\footnote{https://blockchair.com/api/docs} blockchain explorers through their web APIs. These filters can be useful, for example, to retrieve all internal transactions associated with a specific external transaction directly from their transaction cache. Finally, a fourth filter provides the convenience of supplying basic blockchain transaction information inside offline CSV files. Developing filters to provide customized functionalities, such as for an analytics service, e.g. Bloxy\footnote{\url{https://bloxy.info/}}, requires specific development and is left to future work.

\textbf{IoC detectors} are tasked with steps 5-7 according to the different orchestrator calls as defined by the user input. In this respect, the \textbf{explorer} component provides a critical supporting service since it leverages a full archive node to retrieve the relevant historical global states. Given the time-consuming process involved, EtherClue can be configured to cache explorer output. The core task of an IoC detector is to call the user-specified \emph{IoC rule} in step 6. In this regard, the EVM-level detector also needs to pass any additional user parameters, e.g. $TYPE.MIN/MAX$ for the integer over/underflow IoC (see Definition \ref{def-evmoverflowioc}).

\textbf{IoC rules} are JavaScript objects called by detectors per block/trace. A block-level object prototype (or ``class'') must implement the method \texttt{consider}, providing the encoding of IoC semantics, which is called for each block retrieved by its corresponding IoC detector. It exposes the following interface: \texttt{function(blockNumber, txs, accounts, blockBefore, blockAfter)}; where the parameters stand for the block number, its vulnerability-relevant transactions, the relevant historical accounts states, and the pre/post blocks. The EVM-level objects are called per transaction trace. Their \texttt{consider} method is exposed as: \texttt{function(address, tx, trace, callback)}; where parameters stand for the contract address, transaction hash and its trace, along with callback function for reporting alert details. For both object types, a number of helper components are also needed\footnote{See example objects for further details at: \url{https://gitlab.com/simonjaquilina/etherclue/-/tree/master/src/detectors}}. 

\textbf{Context tracker.} This component is used by the EVM-level IoC detector. Whenever a message call-generating instruction is encountered, e.g. \texttt{CALL} and \texttt{STATICCALL}, it ensures that IoC detection is disabled until a \texttt{RETURN} or \texttt{REVERT} appears, otherwise this would create a conflict to the expressions defined over code locations.

\textbf{Modes of operation.} The entire investigation workflow of Figure \ref{fig:arch} supports three specific modes of operation according to the specific choice of components, or their supported functionality. In \emph{local mode} EtherClue is configured to connect to a local archive node, by both the filter and the explorer components. When operating in \emph{cached mode} we refer to IoC detectors using the local cache (step 5b) whenever possible. Finally, the explorer component offers a \emph{custom tracer mode} to the EVM IoC detector. In this mode, \texttt{TxTrace()} accepts a \texttt{tracer} JavaScript object, which is used as input to the \texttt{debug\_traceTransaction} RPC endpoint to pre-filter the instruction trace according to the chosen EVM-level IoC rule, and thus reduce the size of the trace transferred from the archive node.

\textbf{The command-line interface.} The following is an example EtherClue invocation using its command line interface:\\\\
{\footnotesize\texttt{node EtherClue -c "./db" -t "5482796-5488697[blockchair+fs][web3][evm[geth]]" \\-p address=0xc5d105e63711398af9bbff092d4b6769c82f793d \\-f blockchair[from=5482796,to=5488697,fs=["batchTransfer(address[],uint256)"]] \\-e web3[provider=https://api.archivenode.io/XXXXXXXXXXXXXXXX/erigon] \\-d evm[pattern=GethOverflowIoCDetector]}}\\\\
where the \texttt{-t} tag switch identifies the name of the output log, which in this example reflects the ensuing user configuration; the \texttt{-p} shared parameter switch is used for key/value pairs to be made available to all EtherClue components, such as contract addresses; the \texttt{-f} filter switch identifies the filter, with further component-specific parameters inside the square brackets\texttt{[]}. In this example  the \texttt{blockchair} filter is chosen, with the \texttt{from}/\texttt{to} parameters defining the block range to be considered. Moreover, the \texttt{fs} function selector parameter defines a further filtering option based on the target public function name. The \texttt{-e} explorer switch identifies the web3 explorer; the only one developed so far, and with its component-specific parameter \texttt{provider} specifying the  \texttt{archivenode.io} API server. Finally, the \texttt{-d} detector switch specifies the use of the EVM-level IoC detector, along with the corresponding IoC Rule specified by the \texttt{pattern} parameter. Vulnerable contract code locations are supplied by an associated JSON file. The \texttt{-c} switch can be used to enable result caching which expects the name of the directory where the cache files will be stored.

\section{Evaluation}
\label{sec:eval}

A number of diverse experiments were carried out to provide further insight into EtherClue's IoC effectiveness and performance and are presented in the following paragraphs.

\subsection{Experimental Setup} 

First, we employed a mix of synthetic contracts embedding common vulnerabilities and smart contracts from Ethereum mainnet. In this manner, we strike a balance between control over blockchain content and realism. The former were specifically created to maximise variation in blockchain content, thereby enabling a more comprehensive insight into EtherClue than what is possible with exiting mainnet content. The latter provides an opportunity to demonstrate practical deployment. In the former setup, we connect EtherClue to a private \texttt{geth 1.9.21-stable} node whose blockchain is populated with multiple benign and exploit transactions. Both \texttt{geth} and EtherClue are hosted on an Intel Core i5 1.3 GHz with 4GB of RAM machine. For the mainnet setup, we connected EtherClue to both Etherscan and ArchiveNode to obtain the full internal transaction list for each external transaction of interest and the EVM traces per external/internal transaction, respectively.

The datasets\footnote{\url{https://gitlab.com/simonjaquilina/etherclue/-/tree/master/datasets}} used for experimentation have been publicly released to facilitate further experimentation with Ethereum forensics. It includes a private blockchain, aiming to emulate real vulnerable smart contracts and exploit transactions, but purposely aiming to maximise content variation, along with labelled (manually verified) attack transactions for the Ethereum mainnet case studies.

The synthetic smart contracts are: \emph{Underflow} - \texttt{TargetUnderflow} and \texttt{DelayedUnderflow} that include underflow vulnerabilities that directly or indirectly (i.e. overflow impact on control-flow) affect contract storage respectively; \emph{Overflow} -\texttt{SimulationBECToken} which embeds the BECToken overflow vulnerability\footnote{\url{https://peckshield.medium.com/alert-new-batchoverflow-bug-in-multiple-erc20-smart-contracts-cve-2018-10299-511067db6536}}; \emph{Reentrancy} - \texttt{Bank} and \texttt{ProductVote} that insecurely allow recursive Ether withdrawal and multiple-voting at the cost of a single vote respectively; \emph{DoS} - \texttt{SimulationKotET} which embeds the King-of-the-Ether-Throne (KotET)\footnote{\url{https://www.kingoftheether.com/postmortem.html}} DoS vulnerability. The mainnet contracts are: BECToken (Overflow), KotET (Dos), The DAO\footnote{\url{https://www.coindesk.com/understanding-dao-hack-journalists}} (Reentrancy), SpankPay\footnote{\url{https://medium.com/spankchain/we-got-spanked-what-we-know-so-far-d5ed3a0f38fe}} (Reentrancy), PrivateBank\footnote{\url{https://www.reddit.com/r/ethdev/comments/7x5rwr/tricked by a honeypot contract or beaten by/}} (Reentrancy) and Ammbr\footnote{\url{https://medium.com/coinmonks/an-inspection-on-ammbr-amr-bug-a53b4050d52}, \url{https://4hou.win/wordpress/?p=21704}} (Overflow).

\subsection{Results} 

\noindent\textbf{Exploit transaction detection.} Table \ref{tbl:detectionrate} illustrates the detection results of exploit transactions targeting the synthetic contracts at both the EVM and block levels. For each vulnerability, a session of transactions was submitted, consisting of both benign and exploit transactions. Benign transactions comprise vulnerable smart contract deployment and subsequently bring them to a state that renders them appealing for exploitation, e.g. setting up products and votes in \texttt{ProductVote} or the highest bidder in \texttt{SimulationKotET}. Further transactions are sent to the server as background noise in parallel to exploit transactions, enabling a realistic block mining process. The case of \texttt{TargetUnderflow} is quite particular. It represents the case where an underflow results in a storage variable that is propagated to other variables in the same storage, with all transactions ending up being exploits. As such, the benign transaction count varies per smart contract.

Notably, no false positives (FP) were registered in either case, while the EVM-level returned fully accurate detection in all cases i.e. the True Positives (TP) equaled the actual number of exploit transactions, and there were no FP or false negatives (FN). Nevertheless, the block-level missed out on four malicious transactions targeting \texttt{SimulationBECToken}. Three of them were missed since the transactions were internal ones, and their information is not retained at the block level (i.e. by $\beta$) that is investigated. The remaining transaction was missed since the affected storage variable was overridden by a subsequent update in the same block, occluding the attack side-effect in question. It is noteworthy to mention that these limitations were exposed by our approach aiming to maximise variation of blockchain content rather than limiting experimentation solely to currently available mainnet transactions. 

\begin{table}[t]
\captionsetup{justification=centering}
\caption{Exploit transaction detection at the EVM and Block-level IoCs. \\$\dagger$-Over/underflow $\ddagger$-DoS $\star$-Reentrancy}
\footnotesize
\centering
\begin{tabular}{|l|c|c|c|c|c|c|c|c|c|}
\hline
\textbf{Contract Account} & \textbf{TXs} &  \textbf{TP} & \textbf{FP} & \textbf{FN} & \textbf{Blocks} & \textbf{TP}& \textbf{FP} & \textbf{FN}\\
 & \textbf{No.}  &  \textbf{EVM} & \textbf{EVM} & \textbf{EVM} &  \textbf{No.} & \textbf{Block} & \textbf{Block}  & \textbf{Block}\\
\hline
\texttt{Bank$\star$} & 
82 &
6 &
0 &
0 &
42 & 
3 &
0 &
0\\

\hline
\texttt{DelayedUnderflow$\dagger$} & 
23 &
11 &
0 &
0 &
22 & 
11 &
0 &
0\\
\hline
\texttt{ProductVote$\star$} & 
115 &
20 &
0 &
0 &
38 & 
15 &
0 &
0 \\

\hline
\texttt{SimulationBECToken$\dagger$} & 
56 &
12 &
0 &
0 &
25 & 
4 &
0 &
4 \\

\hline
\texttt{SimulationKotET$\ddagger$} & 
151 &
4 &
0 &
0 &
24 & 
4 &
0 &
0\\

\hline
\texttt{TargetUnderflow$\dagger$} & 
20 &
20 &
0 &
0 &
20 & 
20 &
0 &
0\\

\hline
\end{tabular}
\label{tbl:detectionrate}
\end{table}

 \noindent\textbf{Performance.} We measured EtherClue's performance in terms of investigation workflow duration for gradual increases in contract size. The contract size varies in terms of instruction counts, bound to affect EVM-level IoC detectors due to an increase in small-steps, and storage size, bound to affect block-level IoC detectors when fetching contract storage. For the former, we modified \texttt{DelayedUnderflow.sol} with additional \texttt{ADD} instructions (serving as a \emph{NOP}) until we reached the 2 million gas unit utilization mark, far exceeding typical transaction trace size (including the internal calls). For the latter, we took a similar approach with \texttt{TargetUnderflow.sol}, this time by making use of an \texttt{mapping(uint=>uint)} storage variable, conveniently allowing one of its functions to add a key/value pair at each invocation. In both cases, we do not factor in the network latency portions as part of the processing times since they are the same for both the EVM and block-level IoC detectors. Moreover, we did not apply any parallelisation or further optimisations in EtherClue's implementation, leaving this task for future research. In fact, the lack of parallelisation, illustrates the efficacy of our proposed solution since the results can be linearly improved, in terms of time, using more cores.

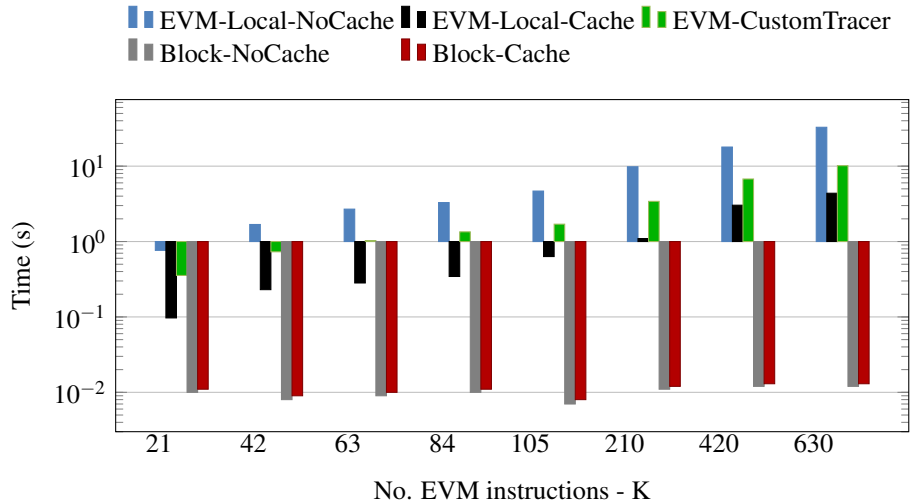
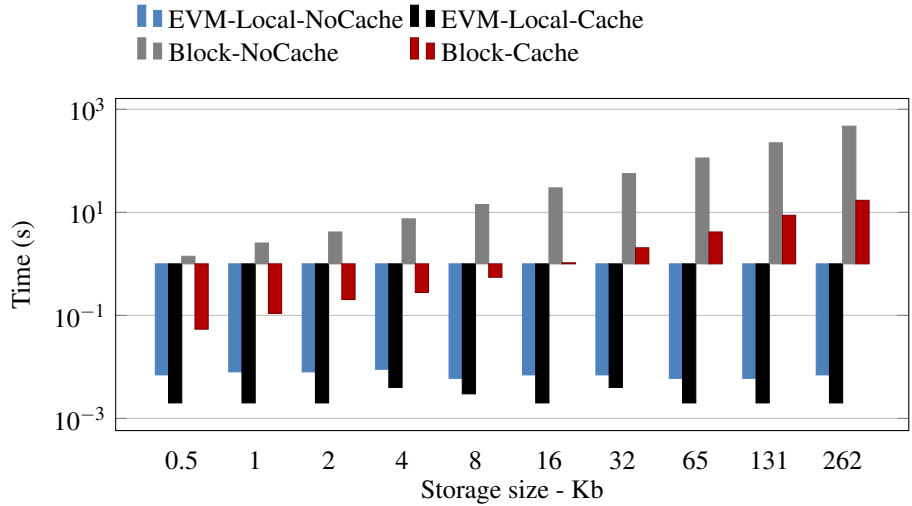
\begin{figure*}[h!]
\centering
\begin{subfigure}[t]{\textwidth}
\begin{tikzpicture}
    \begin{axis}[
        width  = \linewidth,
        height = 6cm,
        ymode=log,
        major x tick style = transparent,
        x tick label style={anchor=east},
        ybar=0pt,
        bar width=4pt,
        ymajorgrids = true,
        ylabel = {Time (s)},
        xlabel = {No. EVM instructions - K},
        symbolic x coords={21, 42, 63, 84, 105, 210, 420, 630},
        xtick = data,
        scaled y ticks = false,
        ymin=0,
        legend columns=3,
        legend cell align=left,
        legend style={
        draw=none,
                at={(0.5,1.3)}, anchor=north,
                column sep=0.1ex
        }
    ]
        \addplot[style={bblue,fill=bblue,mark=none}]
            coordinates {(21, .764) (42,1.684) (63,2.691) (84,3.288) (105,4.673) (210,9.812) (420,17.926) (630,32.789)};

        \addplot[style={black,fill=black,mark=none}]
             coordinates {(21, .097) (42,.230) (63,.282) (84,.344) (105,.632) (210,1.096) (420,3.035) (630,4.374)};

        \addplot[style={ggreen,fill=black!30!green,mark=none}]
             coordinates {(21, .355) (42,.731) (63,1.031) (84,1.344) (105,1.700) (210,3.412) (420,6.755) (630,10.150)};

        \addplot[style={gray,fill=black!50,mark=none}]
             coordinates {(21, .010) (42,.008) (63,.009) (84,.010) (105,.007) (210,.011) (420,.012) (630,.012)};

        \addplot[style={mmaroon,fill=black!30!red,mark=none}]
             coordinates {(21, .011) (42,.009) (63,.010) (84,.011) (105,.008) (210,.012) (420,.013) (630,.013)};

        \legend{EVM-Local-NoCache,EVM-Local-Cache,EVM-CustomTracer,Block-NoCache,Block-Cache}
    \end{axis}
\end{tikzpicture}
\caption{Performance analysis for a gradual increase in contract size in terms of instructions.}
\end{subfigure}

\begin{subfigure}[t]{\textwidth}
\begin{tikzpicture}
    \begin{axis}[
        width  = \linewidth,
        height = 6cm,
        ymode=log,
        major x tick style = transparent,
        ybar=0,
        bar width=5pt,
        ymajorgrids = true,
        ylabel = {Time (s)},
        xlabel = {Storage size - Kb},
        symbolic x coords={0.5, 1, 2, 4, 8, 16, 32, 65, 131, 262},
        xtick = data,
        scaled y ticks = false,
        ymin=0,
        legend cell align=left,
        legend columns=2,
        legend style={
        draw=none,
                at={(0.35,1.3)}, anchor=north,
                column sep=0.1ex
        }
    ]

        \addplot[style={bblue,fill=bblue,mark=none}]
             coordinates {(0.5, 0.007) (1, 0.008) (2,0.008) (4,0.009) (8,0.006) (16,0.007) (32,0.007) (65,0.006) (131,0.006) (262,0.007)};

        \addplot[style={black,fill=black,mark=none}]
             coordinates {(0.5, 0.002) (1, 0.002) (2,0.002) (4,0.004) (8,0.003) (16,0.002) (32,0.004) (65,0.002) (131,0.002) (262,0.002)};

        \addplot[style={gray,fill=black!50,mark=none}]
             coordinates {(0.5, 1.403) (1, 2.537) (2,4.155) (4,7.472) (8,14.160) (16,29.820) (32,56.374) (65,113.34) (131,223.838) (262,470.586)};

        \addplot[style={mmaroon,fill=black!30!red,mark=none}]
             coordinates {(0.5, 0.054) (1,.109) (2,.203) (4,.278) (8,.549) (16,1.048) (32,2.050) (65,4.157) (131,8.769) (262,17.089)};

        \legend{EVM-Local-NoCache,EVM-Local-Cache,Block-NoCache,Block-Cache}
    \end{axis}
\end{tikzpicture}
\caption{Performance analysis for a gradual increase in contract size in terms of storage}
\end{subfigure}

\caption{Performance analysis for a gradual increase in contract size.}
\label{tbl:performance}
\end{figure*}

Figure \ref{tbl:performance} (upper) shows how an increase in the instruction count, as expected, only affects the EVM-level. However, the slow-down can be mitigated by the custom tracer mode. Whenever cached entries can be used, the slow-down can be mitigated even further. In any case, the increase in processing time is linear. On the other hand, Block-level IoCs are slowed down by an increase in contract storage size, as shown in Figure \ref{tbl:performance} (lower). In this case, we did not bother with the custom tracer mode since the processing times for local modes remained constant. Once again, results demonstrate that caching can drastically decrease processing time while the slow-down is linear. Contract size varied in terms of instruction counts, bound to affect EVM-level IoC detectors due to an increase in small-steps and storage size, bound to affect block-level IoC detectors when fetching contract storage. For the former, we modified \texttt{DelayedUnderflow.sol} with additional \texttt{ADD} instructions (serving as a \emph{NOP}) until we reached the 2 million gas unit utilization mark, far exceeding typical transaction trace size (including the internal calls). For the latter, we took a similar approach with \texttt{TargetUnderflow.sol}, this time by making use of an \texttt{mapping(uint=>uint)} storage variable, conveniently allowing one of its functions to add a key/value pair at each invocation. In both cases, we do not factor in the network latency portions as part of the processing time, since they are the same for both the EVM and block-level IoC detectors. Moreover, we did not apply any parallelisation or further optimisations in EtherClue's implementation, leaving this task for future research.

\noindent\textbf{Practicality.} Table \ref{tbl:mainnet} shows detection/duration details of the four mainnet case studies (column 1) conducted over ArchiveNode using the EVM-level IoC detector in custom tracer mode. Based on the result attained using the synthetic accounts, this configuration provides the best option for effectiveness and performance. The EVM-level IoC detector used the Blockchair connector to retrieve all internal transactions resulting from external ones that belong to the chosen block range. Columns 2-4 show the attack date range, followed by the block ranges and their corresponding number of blocks. In all cases, these comprise all the blocks mined in the date range. This range includes just the first day of the attack in each case, except for PrivateBank. In the latter, the related documentation did not specify when the incident occurred. The fifth column shows the full external transaction count included in that block range. Column 6 shows the number of those (external) transactions actually considered to be relevant by the IoC detector. For BECToken, given that the vulnerability is located in the \texttt{batchTransfer()} public function, which is never called internally by the smart contract itself, we investigated all external transactions calling \texttt{batchTransfer()} directly, or indirectly via an internal transaction. As for Ammbr, a similar filter was not applicable so we considered all transactions targeting the vulnerable smart contract. In the case of KotET, all transactions targeting its address are considered, since any of them could result in an internal call that reverts unexpectedly. For TheDAO, SpankPay and PrivateBank, the IoC detector considers all external transactions, targeting any smart contract address, with at least one internal transaction targeting their own address.

\begin{table}[t]
\caption{Ethereum mainnet case studies (EVM-level detection).\\$\dagger$-Overflow $\ddagger$-DoS $\star$-Reentrancy}
\footnotesize
\centering
\resizebox{\textwidth}{!}{%
\begin{tabular}{|l|c|c|c|c|c|c|c|c|}
\hline
\textbf{Contract} & \textbf{Date} &\textbf{Block} &\textbf{Blocks} & \textbf{TX} & \textbf{TX}&\textbf{TP}&\textbf{FP}&\textbf{Duration}\\

 & & \textbf{Range} & \textbf{No.} & \textbf{Full} & \textbf{Relevant} & & & \textbf{(s)} \\

\hline
$\text{BECToken}^{\dagger}$ & 
 22/04/18 &
5482796-5488697 &
5902 &
748858 &
17&
1&
0&
113 \\

\hline
$\text{Ammbr}^{\dagger}$ & 
 07/07/18 &
5918554-5924373 &
5820 &
478209 &
20&
1&
0&
9 \\

\hline
$\text{KotET}^{\ddagger}$ & 
07/02/16&
964910-969996 &
5087 &
14264 &
9&
2 &
0&
3 \\

\hline
$\text{TheDAO}^{\star}$ &
17/06/16&
1717562-1723410&
5849 &
66558 &
5666&
448 &
0&
1838 \\

\hline
$\text{SpankPay}^{\star}$ &
07/10/18&
6467090-6473330&
6241 &
497767 &
201&
7 &
0&
172 \\

\hline

$\text{PrivateBank}^{\star}$ &
30/01/18&
5000001-6000000&
1000000 &
125256429 &
5 &
2 &
0 &
3 \\

&
-20/07/18&
&
&
&
&
&
&
\\

\hline

\end{tabular}
}
\label{tbl:mainnet}
\end{table}

{The results confirm the efficacy of EVM-level detection in the case of synthetic contracts. TP and FP were computed through manual verification of all flagged transactions using EtherScan\footnote{https://etherscan.io/}. In the case of BECtoken and Ammbr, only one transaction was flagged in each case, coinciding with the transaction hashes found in their corresponding investigation reports. For KotET, we checked that the flagged transactions had errors (reverted with an out-of-gas exception) but were still completed. As for the TheDao and SpankPay we checked that the detected transactions had multiple internal message calls to TheDarkDAO (attacker's source account). As seen in the state-of-the-art \cite{zhang2020txspector}, in the lack of appropriate ground truth, false negatives could no be accounted for in a precise manner. However, for TheDAO, SpankPay and PrivateBank, we confirmed that EtherClue was able to detect all exploit transactions detected by Sereum \cite{rodler2018sereum}. Duration times include the entire period as of when the EtherClue orchestrator is invoked with the given block range until all transaction processing terminate. Even though EVM-level processing can end up being time-consuming due to large traces, results show that the investigation duration times remain practical throughout the case studies. As already discussed, the time taken to investigate a single transaction depends on contract sizes, both in terms of instruction counts and storage sizes (see Figure \ref{tbl:performance}). 

Moreover, the total investigation time is impacted by the total number of transactions considered relevant for the case in hand. Taking the case of the TheDAO, which is by far the most expensive among the four case studies, a non-parallel implementation of EtherClue can process a block in 0.3s. This is already faster than the rate at which a new block is currently mined in mainnet, falling in the 10-20s range\footnote{see https://etherscan.io/blocks}, and therefore implying that EtherClue can even be deployed as a real-time detector of exploit transactions.

\noindent\textbf{Comparison with state of the art}. As described in Section \ref{sec:introduction}, EtherClue's scope complements smart contract-centric vulnerability detection tools with a transaction-centric one intended for incident response. In this respect, TxSpector \cite{zhang2020txspector}, Sereum \cite{rodler2018sereum} and DEFIER \cite{suevil} fall into this category, presenting candidate comparison targets. However, one of the main issues of this research topic is that it lacks a standardised framework to leverage a sound comparison between the state of the art methods. The main implications, also mentioned in the literature \cite{zhang2020txspector}, are: the lack of comprehensive ground truth beyond the transaction hashes detected and manually followed up by individual efforts and the availability of openly available rules/analysis routines.

Nevertheless, given that TxSpector's rulesets are openly available, while the transactions detected by Sereum are available in a properly labelled way, we are in a position to compare EtherClue's EVM-level IoCs with TxSpector based on ruleset definition and to cross-check the detected reentrancy exploit transactions with Sereum's. No form of comparisons with DEFIER is yet possible since, firstly, DEFIER requires a set of seed transactions to be made available, something that EtherClue does not assume, while the notion of an attack transaction in DEFIER goes beyond strictly that of exploit transactions to also include the entire attack kill-chain. Moreover, block-level IoC comparison is not possible since none of the existing tools provide it. We expand further on the requirements for a standardised framework for sound and fair comparison in the ensuing discussion (Section \ref{sec:discussion}).

We start by comparing ruleset definition between EtherClue and TxSpector. Therefore, for EtherClue we consider its IoC rules and for TxSpector, its detection rules. Taking the \emph{DoS with (unexpected) revert} as a case study, Listing \ref{lst:txspectordos} shows TxSpector's corresponding detection rule. Its detection engine is implemented using the \texttt{Souffl\'e} high-performance Datalog query tool. Control and data flow dependencies, such as opcode sequences and information flow, are extracted from transactions traces and populate a knowledge base, in turn, queried using relational logic rules. This declarative approach is a strong point for TxSpector; however, it is clear from this example that this tool is intended primarily for users knowledgeable of program analysis. Ultimately, this is the same approach used for static code analysis but adapted to work on runtime traces. In Listing \ref{lst:txspectordos}, line 7 extracts all \texttt{CALL} opcodes, while line 8 checks whether their return values are eventually used by a \texttt{JUMPI} opcode (a conditional jump). Yet, the \texttt{JUMPI} opcode may not immediately follow its corresponding \texttt{CALL}, nor act directly on the returned value itself. Therefore, the rule on lines 1-4 is required to capture this additional complication, using further control/data flow dependencies.
\begin{lstlisting}[label={lst:txspectordos}, caption={TxSpector detection rule for DoS.}]
jumpiDep(jumpiIdx, jumpiDepth, depIdx, depVal) :-
    op_JUMPI(_, _, jumpiCond, jumpiIdx, jumpiDepth, _),
    jumpiIdx > depIdx,
    depends(jumpiCond, depVal).

UncheckedCall(args) :-
    op_CALL(_, _, _, callRet, callIdx, 1, _),
    !jumpiDep(jumpiIdx, 1, callIdx, callRet).
\end{lstlisting}

In EtherClue's case, given IoCs are defined as attack side-effects, it only requires reasoning about individual EVM-level steps. Furthermore, EtherClue is meant to complement vulnerability detectors and so can avail from known vulnerability locations inside the compiled EVM code. Starting with Definition \ref{def-dos}, the IoC rule verifies whether: i) the current opcode is a \texttt{CALL}, ii) it falls within the vulnerability location, and iii) the return value is 0. These three checks are simpler and more elegant than TxSpector's detection rule (Listing 6), which is burdened with ascertaining whether the EVM trace being analyzed corresponds to a vulnerability. Therefore, EtherClue's IoC rules do not have to follow the complex control/data flow dependency analysis which impedes TxSpector's rules.

Notably, the IoC's implementation is not restricted to a specific programming paradigm either. The current EtherClue prototype implements IoC rules in JavaScript, an example fragment of which is shown in Listing \ref{lst:ethercluedos}. \texttt{arStack} on line 2 is an instance of the \emph{call context tracker} described in Section \ref{sec:DFIRtool}, and which is used to step through the transaction trace while keeping track of call-depths and activation records corresponding to internal transactions. Recall from Definition \ref{def-actrec} that the activation record includes all runtime information, encompassing stack/memory content, EVM code and the program counter. \texttt{arStack} is therefore central for the IoC rule verification (i.e., the three previously stated checks related to Definition \ref{def-dos}) performed in lines 3, 7 and 9. Finally, the \texttt{callback()} in line 10 flags the transaction as an exploit. This occurs whenever all checks succeed as per the IoC template in Definition \ref{def:evmtemplate}.

\begin{lstlisting}[label={lst:ethercluedos}, language={Solidity}, caption={EtherClue IoC rule for DoS.}]
...
	var previousStep = arStack.active().nthStep(1);
	if(previousStep.op === Op.CALL){
		var stackAfter = step.stack;
		var r = BigInt(stackAfter[stackAfter.length - 1], 16);
		// Check if the result is an IoC for DoS.
		if(r.eq(0)){
			trace.structLogs.forEach(function(step2, i){
				if(step2.pc == 940 || step2.pc == 941){
       				callback(arStack.active().id, tx, previousStep, index, "");
					}
			});
		}
	}
...

\end{lstlisting}
The main observation here is that it should be simpler to implement a series of checks on the pre/post states of EVM instructions rather than defining vulnerabilities using program analysis. Furthermore, EtherClue's IoCs can also be used at the block level, something that TxSpector does not provide. We elaborate further on this point in the upcoming discussion.

Another important aspect that prevents a fair comparison is that, for instance, TxSpector discards transactions not processed according to a timeout threshold. Therefore, the sample set may vary when comparing with other tools such as EtherClue, since the latter analyses all the transactions and is not restricted by performance thresholds. While we believe that the detection tools' performance is relevant, enabling a timeout requires further discussion since individuals may not care about the time analysis if their losses are enormous.

Finally, we further elaborate on the outcomes of TheDAO experiment. According to the original reports, Sereum flagged 2112 transactions whilst TxSpector flagged 2008. EtherClue was able to flag 2208, since it identified 96 exploit transactions that ended up with an exception, and therefore failed to update the blockchain state. According to Sereum documentation, it does not consider non-successful exploit transactions, yet during our analysis, we found out that 7 of such transactions were included in their outcomes, probably due to a bug/error. The latter means that EtherClue achieved the same performance as Sereum (i.e., a total of 2015 transactions if we remove the non-successful ones), which is a tool focused on this specific exploit.

\subsection{Discussion} \label{sec:discussion}
Our research and experimental findings provide some rather interesting insights about blockchain forensics and the corresponding IoCs. In the following paragraphs we discuss their main points.

\noindent\textbf{EVM-level IoCs provide the best option.} EVM-level IoCs can carry a considerably larger computational load as transaction traces get larger. This property of EtherClue can be emphasised for transactions involving multiple internal message calls. On the other hand, this is not an issue for block-level IoCs since contract storage size is independent of the number of executed internal transactions. Nonetheless, one should take into consideration that such IoCs can generalise over an entire exploit class. Moreover, they can specifically flag exploit transactions within their block, and their effectiveness cannot undermined neither by the `unavailable internal transactions' nor by the `occluded side-effect' issues of the block level. Therefore, EVM-level IoC-centric investigations can be more effective. While such IoCs may impose a much larger computational load, our extended experiments illustrate that the cost is tolerable and applicable in real-world scenarios. Beyond a pre-processing step involving transaction tracing, a larger number of analysis steps is also required due to the small-step execution semantics. Yet, the mainnet case studies show that processing times remain practical even for transactions that consume the largest possible amount of gas.

\noindent\textbf{Block-level IoCs can still be of value.} During investigations in which response times are critical, the lightweight nature of IoCs at this level may still render them useful in the form of an `approximate' triage-pass. Albeit not fully accurate, this pass can still help with the quick identification of a sub-set of exploit transactions, especially given that IoCs at this level err on false negatives rather than false positives.

\noindent\textbf{Automated IoC definition/code generation.} The potential value of a triage-pass based on block-level IoCs, however, is undermined by the fact that they are contract-specific, and IoC definition followed by IoC Rule implementation can be time-consuming. Yet, the formal semantics upon which IoCs are defined can provide a solid basis for automation. More precisely, the symbolic execution of a sample exploit transaction could provide the basis for automation by generating all the states associated with successful exploitation and from which to derive IoCs directly in code form. Given that symbolic execution does not require concrete inputs, this enables the generation of IoCs in a generic manner through symbolic inputs. Furthermore, symbolic execution engines for the EVM, e.g. Manticore \cite{mossberg2019manticore}, are already available.

\noindent\textbf{A standardised framework for comparison.} Ultimately, given the limited way we were able to compare EtherClue with existing tools so far, further enhancements require the support of clear comparisons for proper evaluation. Specifically, we encountered substantial difficulties in making use of datasets, at least beyond the limited one provided by Sereum \cite{rodler2018sereum}, due to the lack of properly labelled ground truth and harmonized definitions of what constitutes an exploit transaction, amongst others. Therefore, we believe that studying standardised criteria to measure the quality of the tools with unified benchmarks and metrics is a promising research line in its own right. From our own experience, we propose that such a framework should at least cover: i) Tool categorisation in terms of which modes they can operate in, i.e. contract, block, transaction, or individual EVM instruction levels; ii) In the case of tools supporting a contract-level operation mode, whether they require the source code or whether they can work with just the decompiled bytecode; iii) Make the ruleset/analysis routines available for comparison, at least in object form, but more importantly highlighting the exact scope of their detection; iv) A ground-truth dataset that uses different labels for attack transactions at the different stages of the attack kill-chains, specifically identifying exploit transactions as only those that directly exercise a security bug;  v) Finally, enhance all exploit transaction labels with related block-level, internal message call and instruction-level annotations.

\section{Conclusions}
\label{sec:conclusion}
In this work, we showed that IoCs defined over the side-effects of Ethereum smart contract execution is an effective way to identify exploit transactions. The primary contribution is EtherClue, a DFIR tool that complements vulnerability detectors by post-factum exploit transaction detection. A model that abstracts smart contract execution underpins the method for defining IoCs, and which in turn builds upon a formalism; previously used for symbolic smart contract execution, to fit the needs of digital forensics. Experimentation with common exploit classes demonstrates the effectiveness and real-world practicality of EtherClue compared with other state-of-the-art approaches. Results point towards defining IoCs at the EVM level, rather than the block level since IoCs can generalise over exploit classes and achieve better precision. 

Future work will focus on defining further IoCs for the ever-growing list of exploit classes, along with further experimentation on mainnet. This effort will, however, firstly require a parallel implementation of EtherClue IoC detectors. This optimised implementation will also be useful to experiment further with the idea of applying EtherClue real-time detection, and perhaps even for early detection inside transaction pools. Deriving IoCs automatically from exploit samples is also a possibility given EtherClue's operational semantics foundation. Finally, by integrating EtherClue with data mining tools, e.g. DEFIER, and blockchain analytics, e.g. Bloxy, it will be possible to automate the process of forensic timeline analysis, providing investigators with a complete insight into an incident with minimal manual effort. Moreover, we argue that a standardised framework for tool comparison can be a catalyst for advancing the state of the art in Ethereum forensics and blockchain technologies in general.

\section*{Acknowledgements}

This work was supported by the European Commission under the Horizon 2020 Programme (H2020), as part of the \textit{LOCARD} (\url{https://locard.eu}) (Grant Agreement no. 832735) project.

The content of this article does not reflect the official opinion of the European Union. Responsibility for the information and views expressed therein lies entirely with the authors.

\bibliographystyle{elsarticle-num}
\bibliography{myrefs}

\end{document}